%% file: 15574.tex
\documentclass[traditabstract]{aa}  
\usepackage{psfig,longtable,lscape,amssym,natbib}  
\newcommand{\ltsima} {$\; \buildrel < \over \sim \;$}  
\newcommand{\gtsima} {$\; \buildrel > \over \sim \;$}  
\newcommand{\lta} {\lower.5ex\hbox{\ltsima}}  
\newcommand{\gta} {\lower.5ex\hbox{\gtsima}}  
\newcommand{\Ha} {H$\alpha$}  
\newcommand{\Hb} {H$\beta$}

\newcommand{\ergsHz}{\>{\rm erg}\,{\rm s}^{-1}\,{\rm Hz}^{-1}}

\defcitealias{buttiglione09alias}{Paper~I}
\defcitealias{buttiglione10alias}{Paper~II}
\begin{document}

\title{An optical spectroscopic survey of the 3CR sample of radio galaxies
  with $z<0.3$.}  \subtitle{ III.  Completing the sample\thanks{Based on
    observations made with the Italian Telescopio Nazionale Galileo operated
    on the island of La Palma by the Centro Galileo Galilei of INAF (Istituto
    Nazionale di Astrofisica) at the Spanish Observatorio del Roque del los
    Muchachos of the Instituto de Astrofısica de Canarias.}}

\titlerunning{Optical spectra of 3CR sources} \authorrunning{S. Buttiglione et
  al.}
  
\author{Sara Buttiglione \inst{1,2} \and Alessandro Capetti \inst{3} \and
  Annalisa Celotti \inst{2} \and David J. Axon \inst{4,5} \and Marco Chiaberge
  \inst{6,7} \and F. Duccio Macchetto \inst{6} \and William B. Sparks \inst{6} }
   
\offprints{S. Buttiglione}
     
\institute{INAF, Osservatorio Astronomico di Padova, Vicolo dell'Osservatorio
  5, I-35122 Padova, Italy \email{sara.buttiglione@oapd.inaf.it} \and
  SISSA-ISAS, Via Beirut 2-4, I-34014 Trieste, Italy \and INAF - Osservatorio
  Astronomico di Torino, Strada Osservatorio 20, I-10025 Pino Torinese, Italy
  \and School of Mathematical and Physical Sciences, University of Sussex,
  Falmer, Brighton BN1 9RH, UK \and Department of Physics, Rochester Institute
  of Technology, 85 Lomb Memorial Drive, Rochester, NY 14623, USA \and Space
  Telescope Science Institute, 3700 San Martin Drive, Baltimore, MD 21218,
  U.S.A. \and INAF-Istituto di Radio Astronomia, via P. Gobetti 101, I-40129
  Bologna, Italy }

\date{}

\abstract 
{We present optical nuclear spectra for nine 3CR radio sources obtained with the
Telescopio Nazionale Galileo, that complete our spectroscopic observations
of the sample up to redshifts $<$ 0.3. We measure emission line luminosities
and ratios, and derive a spectroscopic classification for these sources.}  

\maketitle
  
\section{Introduction}
\label{introduction}

The 3CR catalog of radio sources represents a particularly well suited sample
for a study of the physics of radio-loud AGN. Its selection criteria are
unbiased with respect to optical properties and orientation, and it spans a
relatively wide range in redshift and radio power. A vast suite of
observations is already available for this sample, from multi-band HST
imaging, to observations with Chandra, Spitzer and the VLA.

Quite surprisingly, however, the
available optical spectroscopic data for the 3CR sample were sparse and
incomplete. 
To fill this gap, we carried out a homogeneous and complete survey of optical
spectroscopy, targeting the subsample of 113 3CR radio sources
with z$<$0.3, for which we can obtain uniform uninterrupted coverage of the
key spectroscopic optical diagnostics. The observed sources include a
significant number of powerful classical FR~II RG, as well as the more common
(at low redshift) FR~Is \citep{fanaroff74}, spanning four orders of
magnitude in radio luminosity, thus providing a broad representation of the
spectroscopic properties of radio-loud AGN. The data were presented
in \citet{buttiglione09alias} (hereafter Paper I) and discussed  
in \citet{buttiglione10alias} (hereafter Paper II).

However, nine sources of our sample (namely 3C~020, 3C~063, 3C~132, 3C~288,
3C~346, 3C~349, 3C~403.1, 3C~410, 3C~458) could not be observed due to
scheduling problems and time constraints.  Furthermore, the SDSS spectrum of
3C~270 presented in Paper I could not be used for its spectroscopic
characterization, since the fiber was not positioned on the galaxy's nucleus
(Christian Leipski, private communication). In order to reach completeness of
the spectroscopic survey we present the results of new TNG observations
of nine missing sources, while we complemented our data with those obtained
for 3C~270 by \citet{ho97}.

The paper is organized as follows: in Sect. \ref{sect1} we present
the observational procedure and the data reduction, leading to the
measurements of the emission line fluxes (Sect. \ref{sect3}).
In Sect. \ref{sect4} we derive a spectroscopic classification for these
sources, updating the results derived in Paper II.
A brief summary is given in Sect. \ref{summary}.

Throughout, we have used $H_o = 71$ km s$^{-1}$ Mpc$^{-1}$, $\Omega_{\Lambda}
= 0.73$ and $\Omega_m = 0.27$.

\section{Observations and data reduction}
\label{sect1}

The optical spectra of the nine missing 3CR sources were taken with the
Telescopio Nazionale Galileo (TNG), a 3.58 m telescope located on the Roque de
los Muchachos in La Palma Canary Island (Spain). The observations were made
with the DOLORES (Device Optimized for the LOw RESolution) spectrograph.  The
detector used is a 2100x2100 pixels back-illuminated E2V4240, with a pixel
size of 0\farcs252. The observations were carried out in service mode between
September 2008 and July 2009. The chosen 

\begin{landscape}
\begin{figure}[htbp]

\centerline{
\psfig{figure=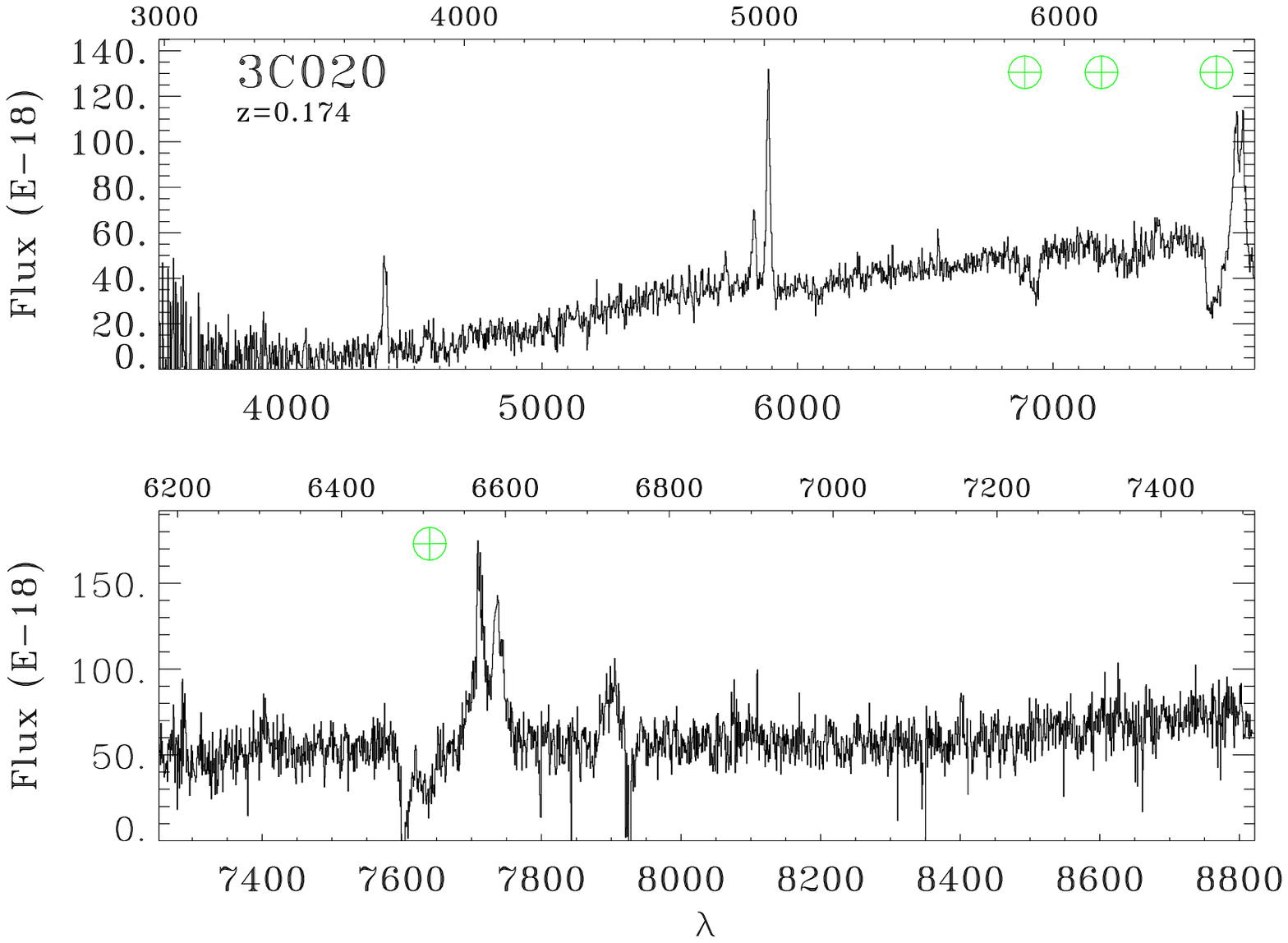,width=0.32\linewidth}
\psfig{figure=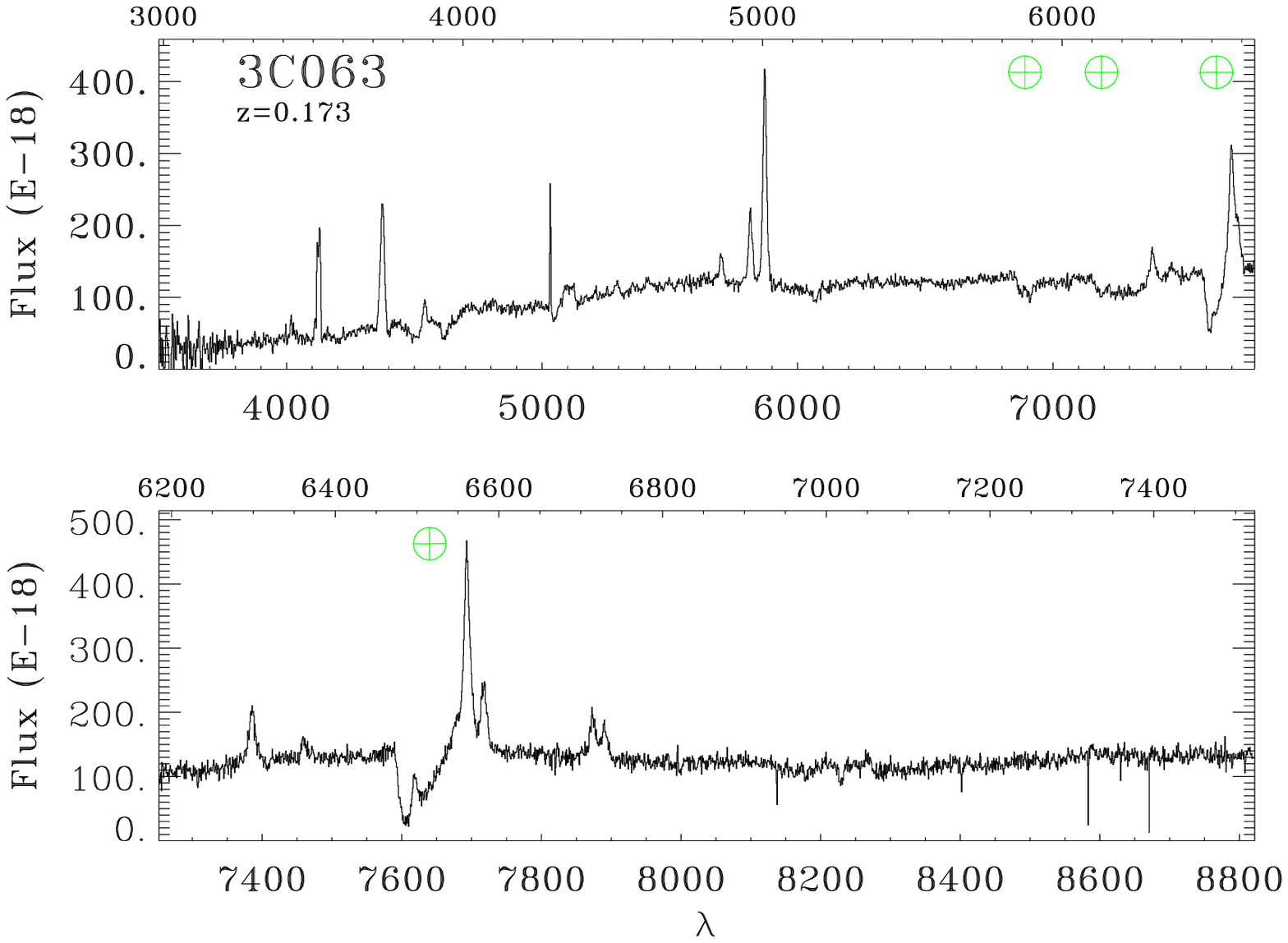,width=0.32\linewidth}
\psfig{figure=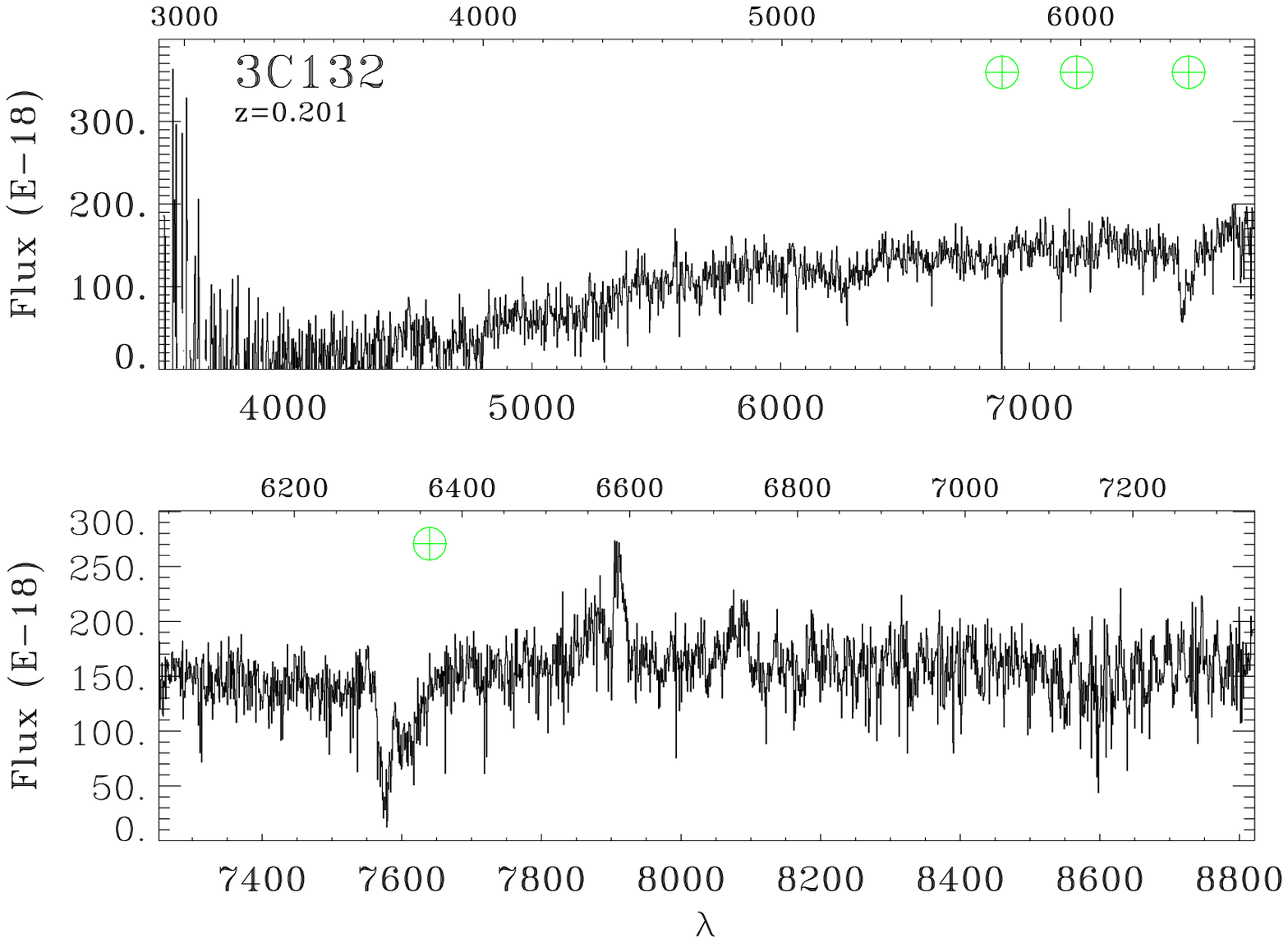,width=0.32\linewidth}}
\centerline{
\psfig{figure=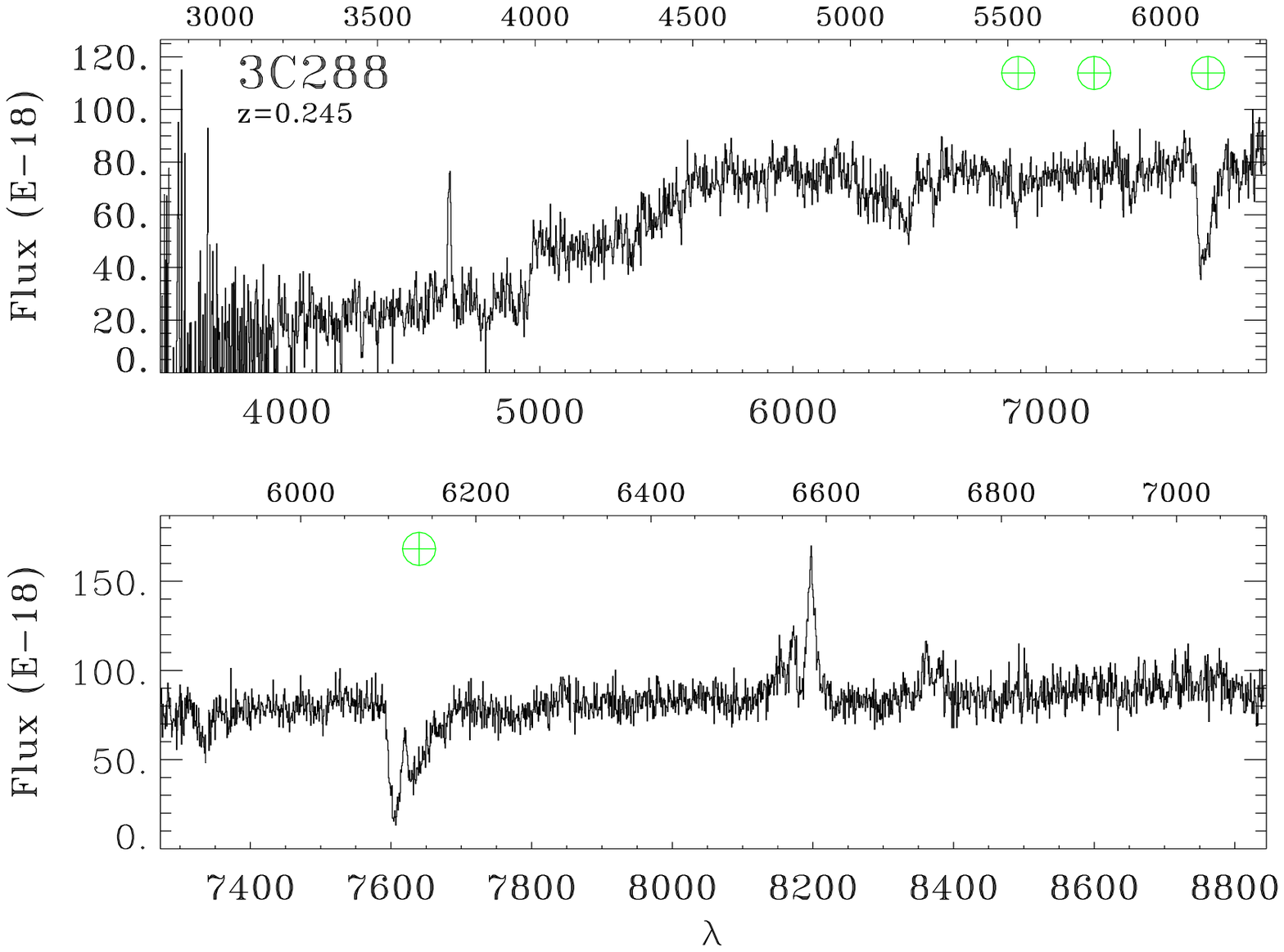,width=0.32\linewidth}
\psfig{figure=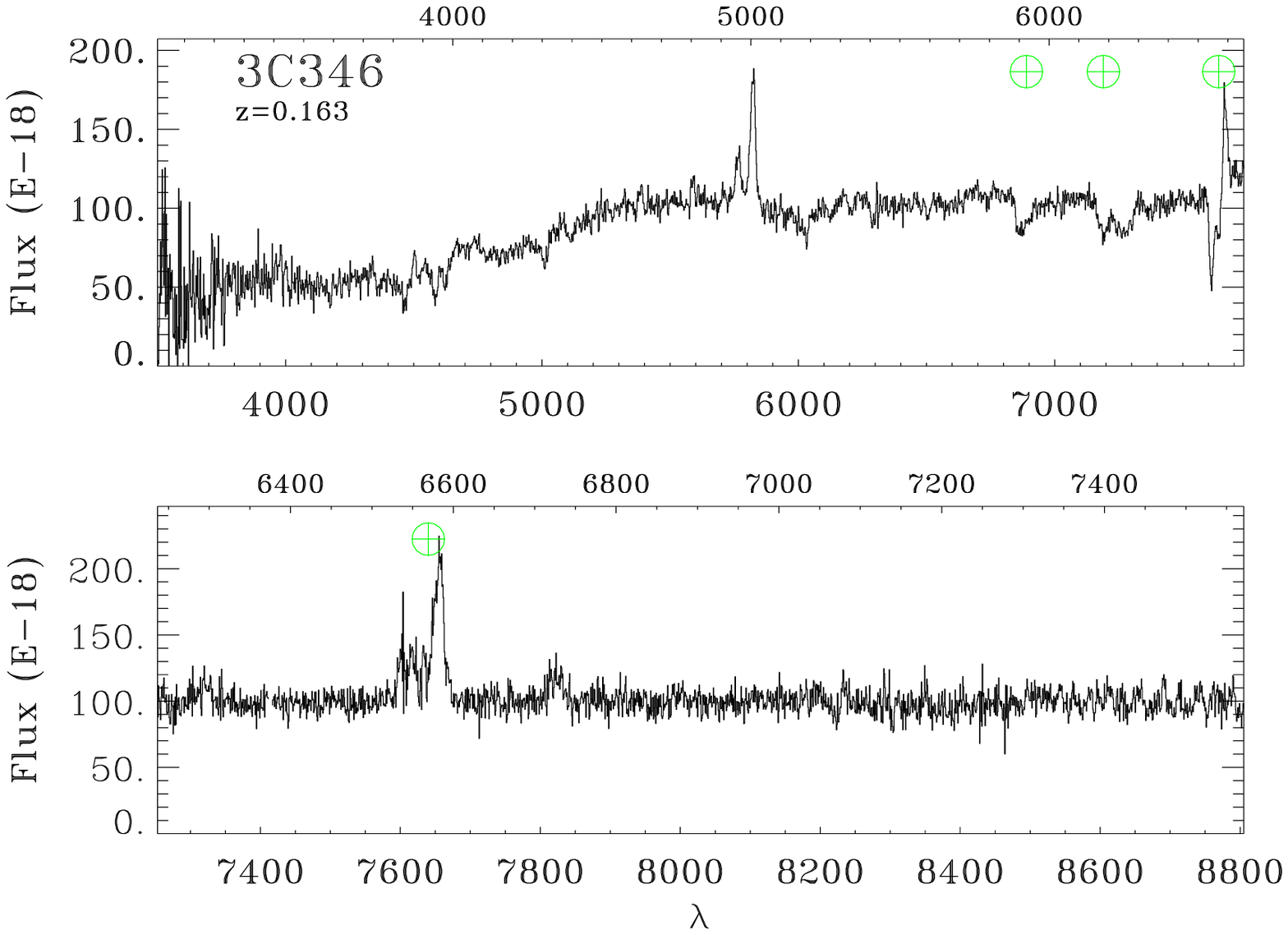,width=0.32\linewidth}
\psfig{figure=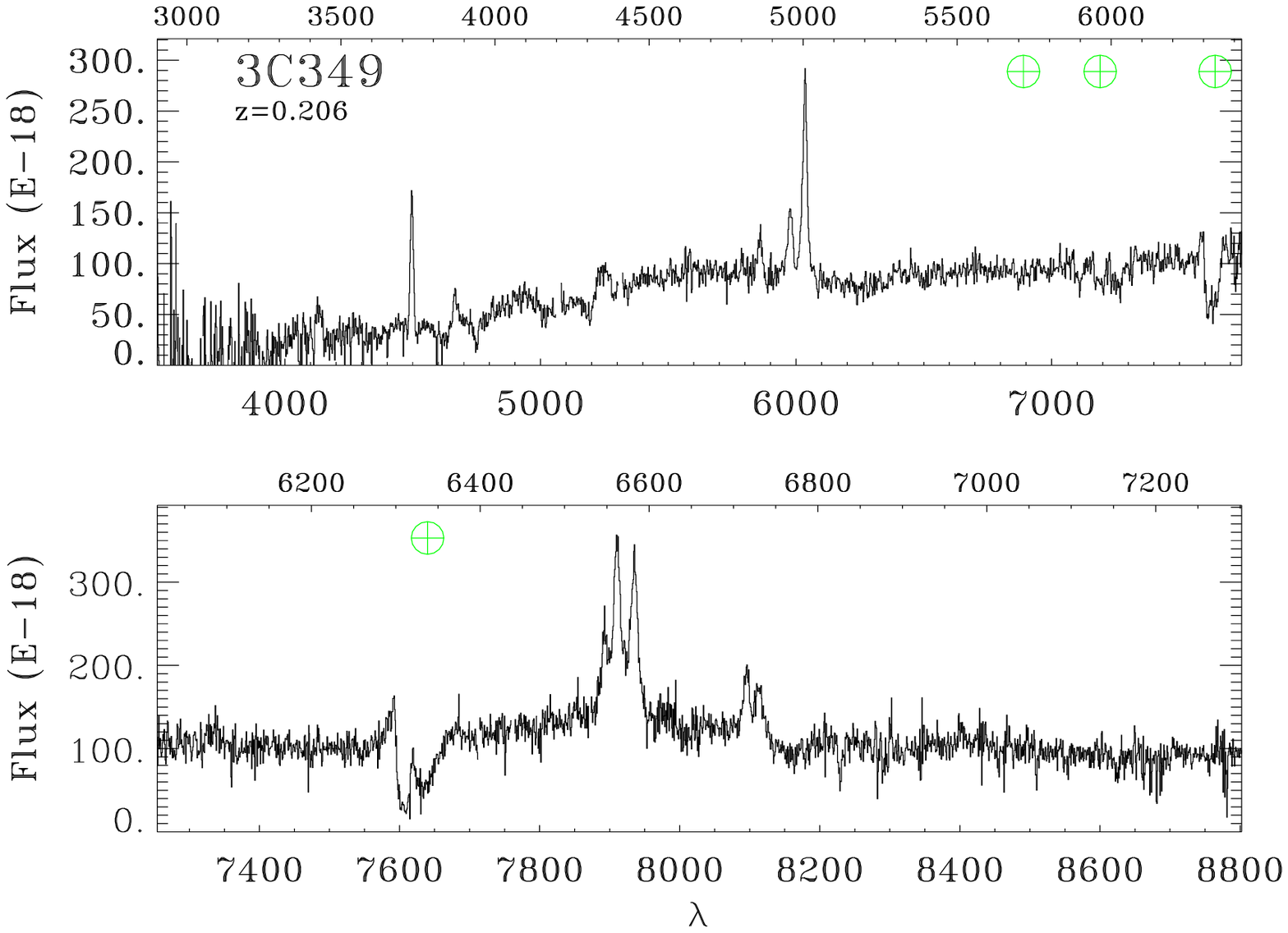,width=0.32\linewidth}}
\centerline{
\psfig{figure=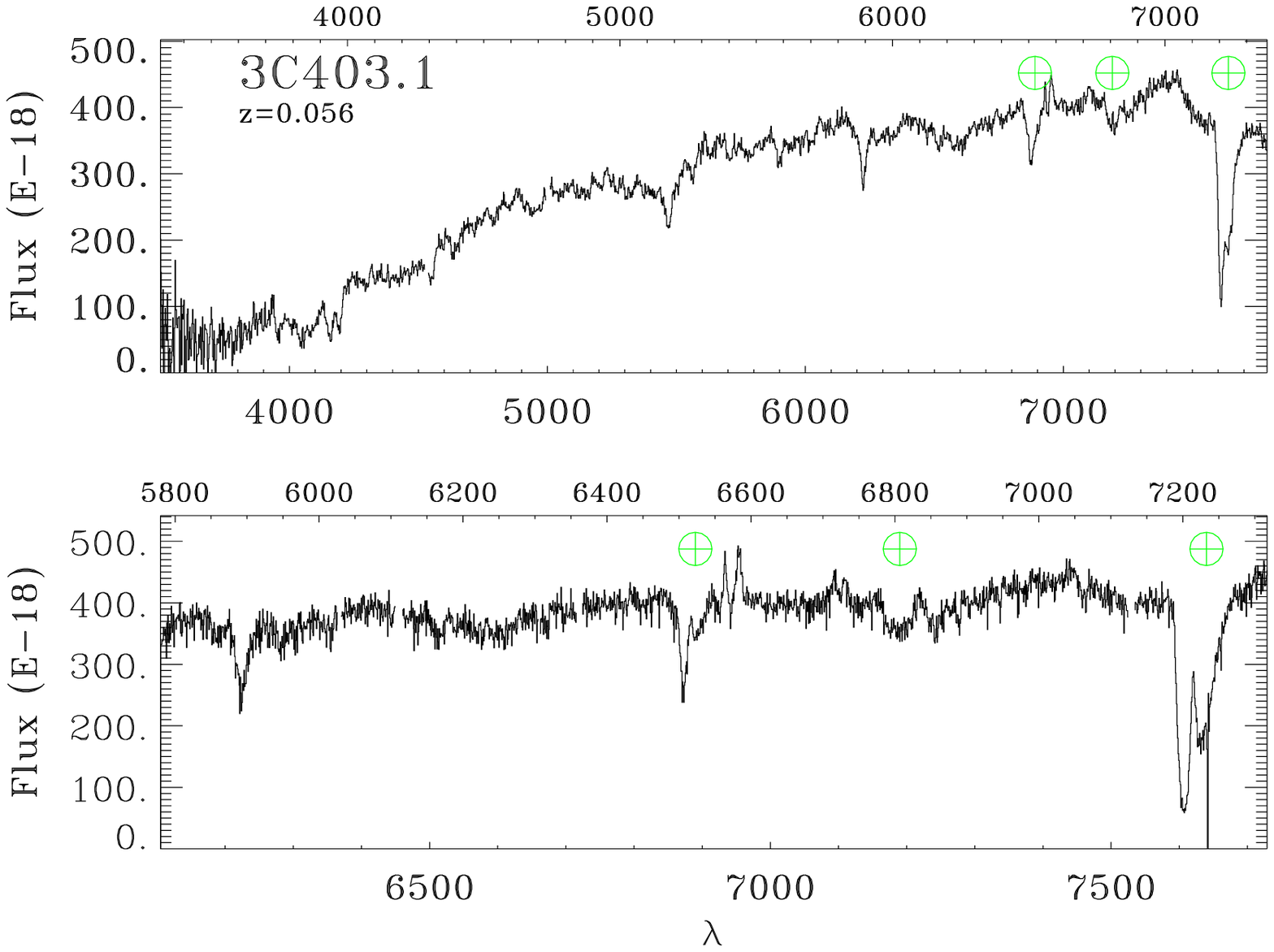,width=0.32\linewidth}
\psfig{figure=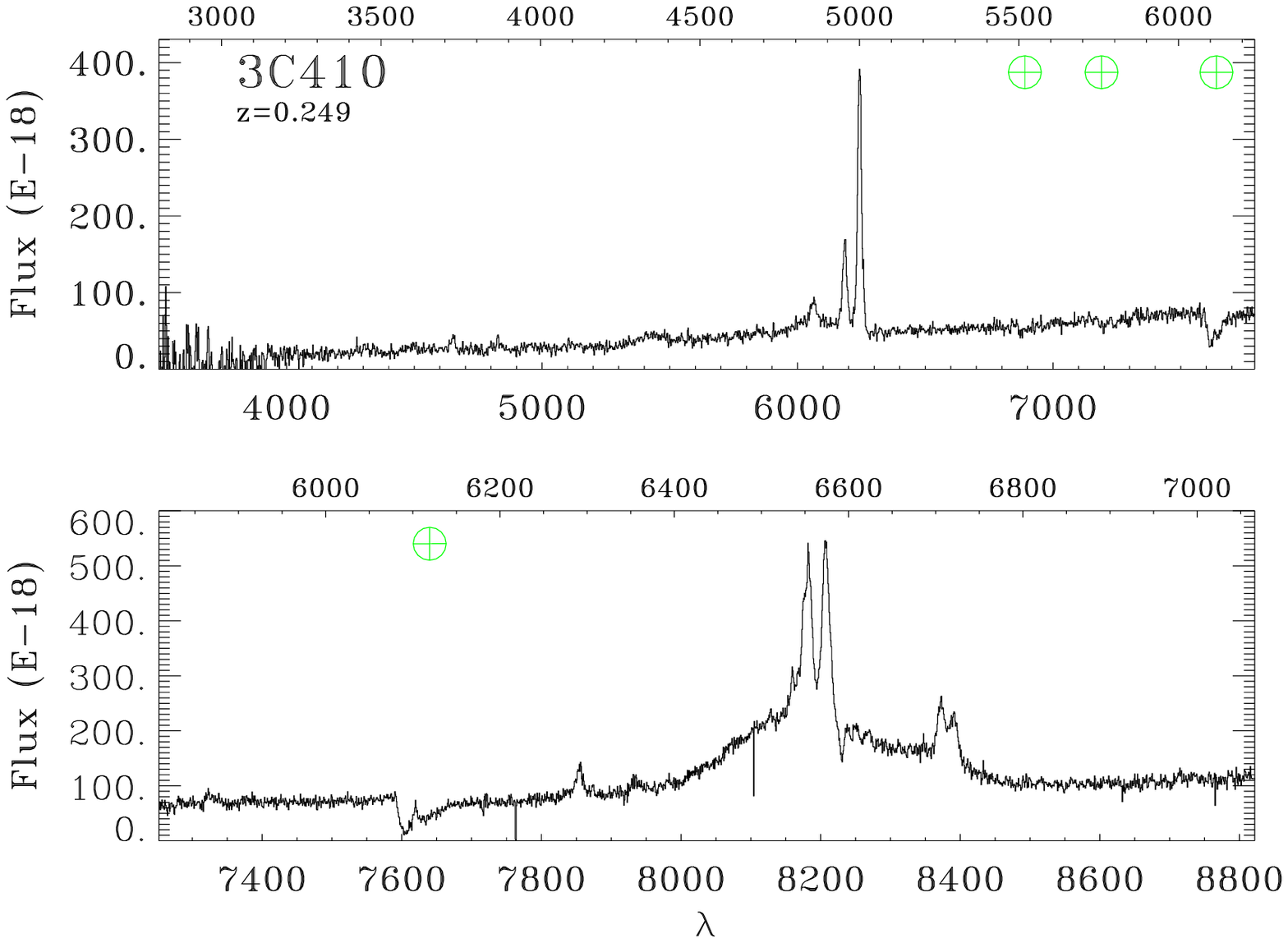,width=0.32\linewidth}
\psfig{figure=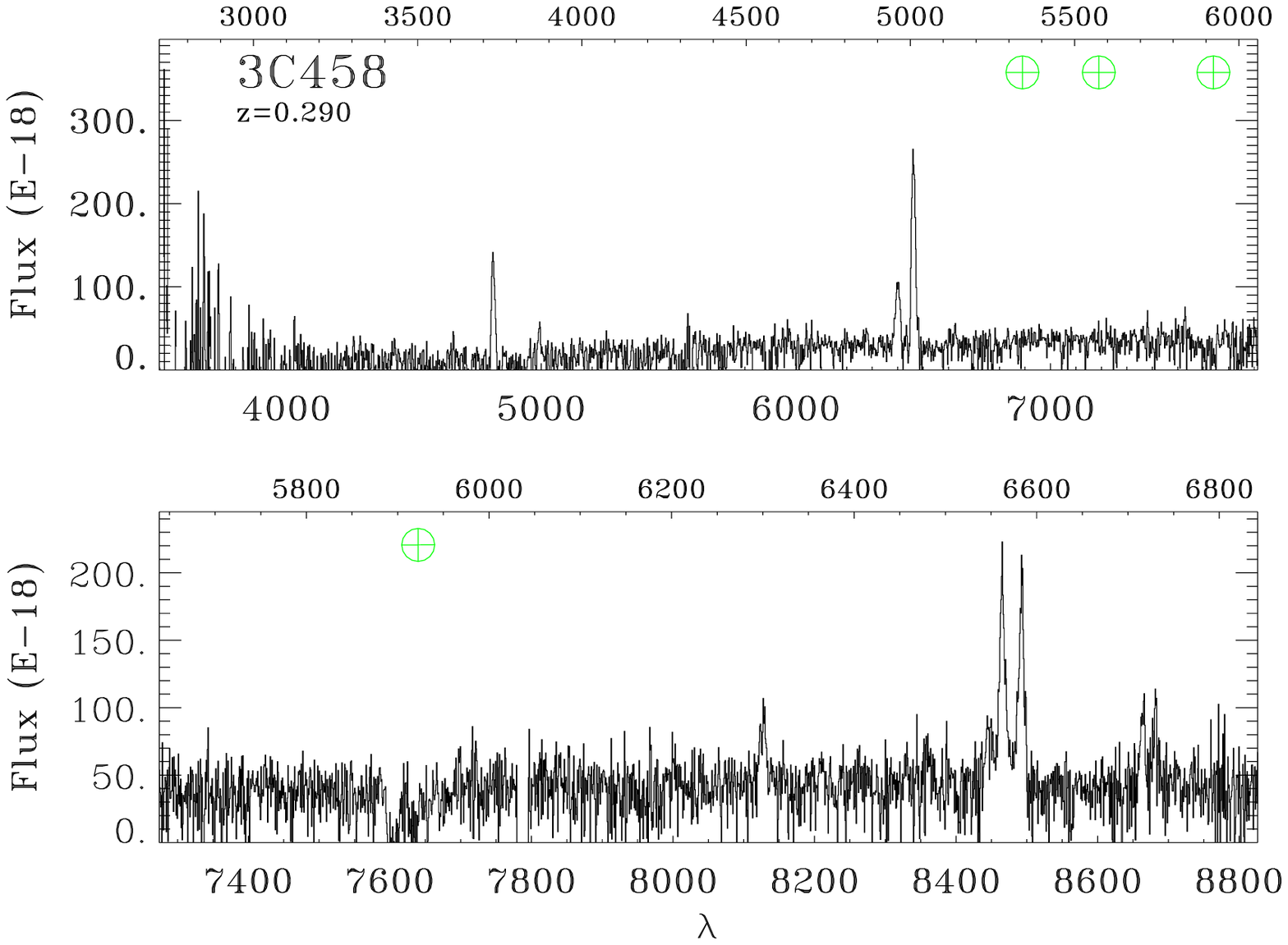,width=0.32\linewidth}}
\caption{\label{spectra} For each source the low (upper panels) 
  and high resolution (bottom panels) spectra are shown. The fluxes are
  in units of erg cm$^{-2}$ s$^{-1}$ \AA$^{-1}$ while the wavelengths are in
  \AA. The lower axes of the spectra show the observed wavelengths
  while the upper axes show the rest frame wavelengths. The three main
  telluric absorption bands are indicated with circled crosses.}
\end{figure}
\end{landscape}

\noindent
long-slit width is 2$\arcsec$ and it
was aligned along the parallactic angle in order to minimize light losses due
to atmospheric dispersion.

For each target we took one (or two) low resolution spectrum with the LR-B
grism ($\sim$3500-7700 \AA) with a resolution of $\sim$20 \AA\ and two high
resolution spectra with the VHR-R (6100-7700 \AA) or VHR-I (7250-8800 \AA)
grisms, depending on redshift, with a resolution of $\sim$5 \AA. 
The exposure times increase with redshift, in order to
compensate for the galaxies' dimming.
Exposures longer than 750 s were divided into two sub-exposures of 500 sec,
obtained moving the target along the slit.  The high resolution spectra have
an exposure time twice the low resolution ones. The combination of the LR-B
and VHR ranges of wavelengths enables us to cover the most prominent emission
lines of the optical spectrum and in particular the key diagnostic lines
H$\beta$, [O~III]$\lambda\lambda$4959,5007, [O~I]$\lambda\lambda$6300,64,
H$\alpha$, [N~II]$\lambda\lambda$6548,84, [S~II]$\lambda\lambda$6716,31. The
high resolution spectra are a sort of {\it zoom} on the H$\alpha$ region with
the aim of resolving the H$\alpha$ from the [N~II] doublet, as well as the two
lines of the [S~II] doublet.  Table \ref{logoss} provides the journal of
observations and the main information on the sources.

The data analysis was performed as describer in
\citetalias{buttiglione09alias}, which should be refer to for further
details. Summarizing the spectra were bias subtracted and flat fielded. When
the spectra were split into two sub-exposures, they were subtracted to remove
the sky background. The residual background was subtracted measuring the
average on each pixel along the dispersion direction in spatial regions
immediately surrounding the source spectrum. The data were then wavelength
calibrated and corrected for optical distortions.  Finally the spectra were
extracted and summed over a region of 2$\arcsec$ along the spatial direction
and flux calibrated using spectrophotometric standard stars, observed
immediately after each target.

The telluric absorption bands were usually left uncorrected except in the few
cases in which an emission line of interest fell into these bands. In these
cases we corrected the atmospheric absorptions using the associated standard
stars as templates.

 Fig. \ref{spectra} shows, for all the observed targets, the low resolution
spectrum (upper image) and the high resolution one (bottom image). The
calibrated spectra are in units of $10^{-18}$ erg cm$^{-2}$ s$^{-1}$
\AA$^{-1}$. The wavelengths (in \AA\ units) are in the observer frame in
the axes below the images while they are in the source frame in the axes above
them. 

\section{Data analysis}
\label{sect3}

We corrected the spectra for
reddening due to the Galaxy \citep{burstein82,burstein84} using the extinction
law of \citet{cardelli89}. The galactic extinction used for each object
was taken from the NASA Extragalactic Database (NED) database and is listed in
Table \ref{bigtable}. We also transform the spectra into rest frame
wavelengths using the value of redshift from NED.

The contribution of stars to our spectra was subtracted using the best fit
single stellar population (SSP) model taken from the \citet{bruzual03} library
out of a grid of 33 single stellar population models, with a Salpeter Initial
Mass Function, formed in an instantaneous burst.  We excluded from the fit the
spectral regions corresponding to emission lines, as well as other regions
affected by telluric absorption, cosmic rays or other impurities. In
  3C~410 the continuum is essentially featureless and it is likely to be
  dominated by non-stellar emission, a characteristic already seen in several
  3CR objects. No starlight subtraction was performed for this object.

By using the {\it specfit} package in IRAF, we then measured line intensities
fitting Gaussian profiles to H$\beta$, [O~III]$\lambda\lambda$4959,5007,
[O~I]$\lambda\lambda$6300,64, H$\alpha$, [N~II]$\lambda\lambda$6548,84, and
[S~II]$\lambda\lambda$6716,31. Some constraints were adopted to reduce the
number of free parameters: we required the widths and the velocity to be the
same for all the lines.  The integrated fluxes of each line were free to vary
except for those with known ratios from atomic physics: i.e. the
[O~I]$\lambda\lambda$6300,64, [O~III]$\lambda\lambda$4959,5007 and
[N~II]$\lambda\lambda$6548,84 doublets. Prominent broad \Ha\ and \Hb\
  lines are visible in the spectrum of 3C~410. We then fit the line emission
  including a broad component. This is well reproduced by a gaussian profile,
  when allowing a small line asymmetry.

Table \ref{bigtable} summarizes the intensities of the main emission lines
(de-reddened for Galactic absorption) relative to the intensity of the narrow
component of H$\alpha$, for which we give flux and luminosity. To each line we
associated its relative error, as a percentage.  We placed upper limits at a
3$\sigma$ level to the undetected, but diagnostically important, emission
lines by measuring the noise level in the regions surrounding the expected
positions of the lines, and adopting as line width the instrumental
resolution. In the case of 3C~346 the telluric correction is not sufficiently
accurate to recover the flux of its \Ha\ line, that falls in a deep
transmission through of a telluric band.  For 3C~410 we also give the flux of
its broad \Ha\ line.

\section{Results}
\label{sect4}

\begin{figure*}[t!]
  \centerline{ \psfig{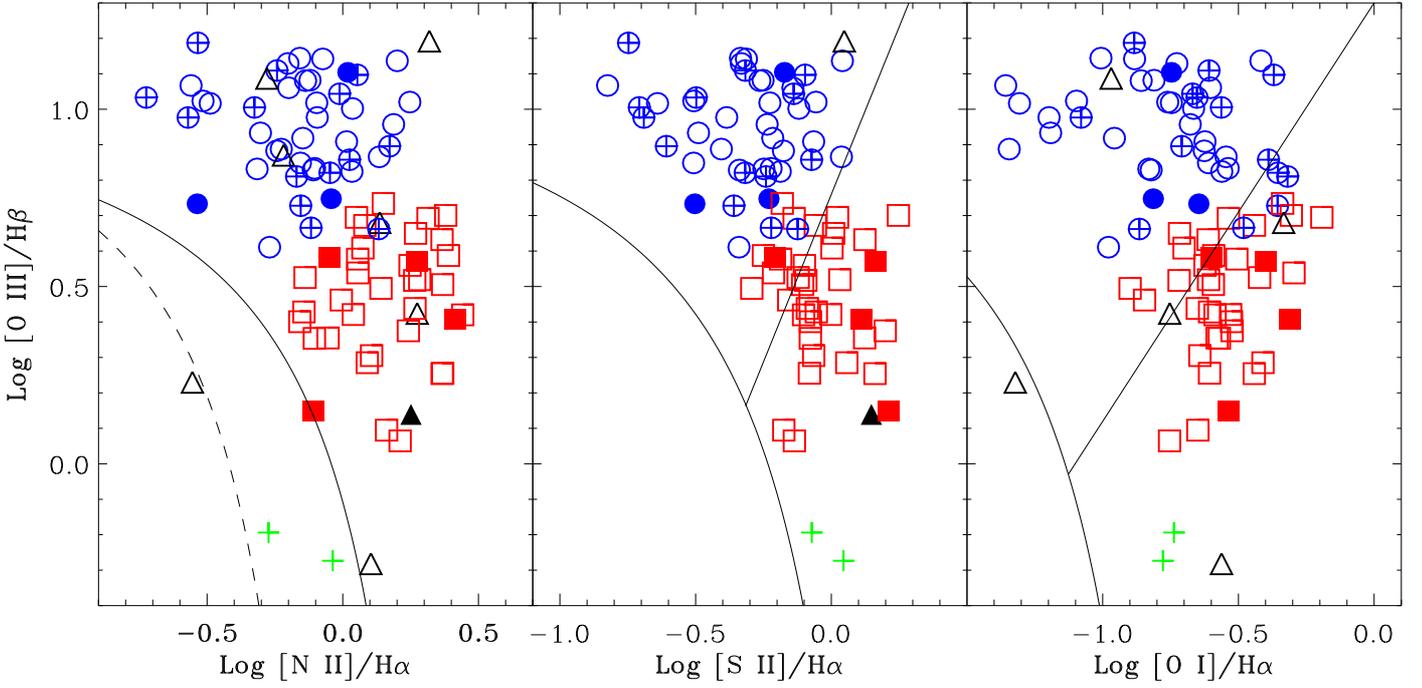}}
  \caption{ Diagnostic diagrams for 3CR sources after the classification into
    HEG (blue circles) and LEG (red squares) made using the Excitation Index.
    Crossed circles are broad line galaxies, green crosses are extremely low
    [O~III]/\Hb\ sources. Black triangles are sources for which the E.I.
    cannot be estimated, as they lack the measurement of one or two 
diagnostic lines. The filled symbols are derived from the new data presented
in this paper.}
  \label{diag-lri}
\end{figure*}

\begin{figure*}[htbp]
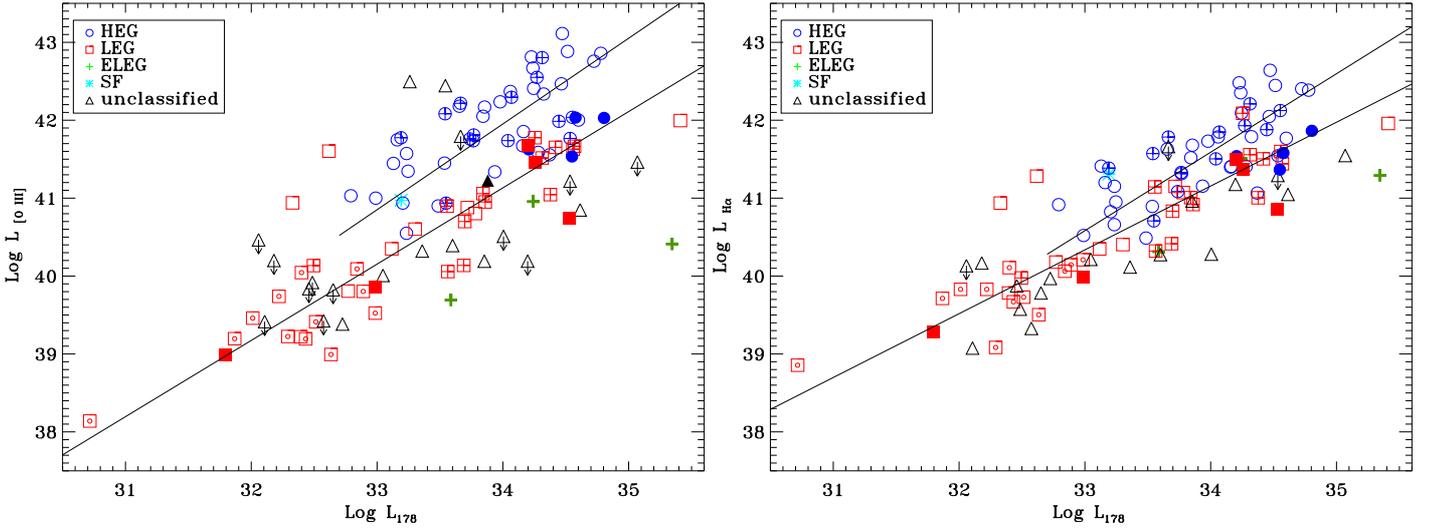

\centerline{
\psfig{figure=15574f3a.epsi,angle=90,width=0.5\linewidth}
\psfig{figure=15574f3b.epsi,angle=90,width=0.5\linewidth}}
\caption{\label{o3re} [O~III] and \Ha\ luminosity [erg s$^{-1}$] (left and
  right panel, respectively) as a function of radio luminosity at 178 MHz [erg
  s$^{-1}$ Hz$^{-1}$]. Blue circles are HEG (crossed circles are BLO), red
  squares are LEG, green pluses are ELEG, the cyan asterisk is the star
  forming galaxy, while the black triangles are spectroscopically unclassified
  galaxies. The two solid lines represent the best linear fit obtained for the
  HEG and LEG sub-populations separately.  When possible, we further mark the
  LEG according to their FR type: crossed squares are FR~II/LEG and dotted
  squares are FR~I/LEG. The filled symbols are derived from the new data
  presented in this paper.}
\end{figure*}

The data quality of the nine spectra considered here is such that we could
measure all diagnostic lines with only three exceptions: [O~I] in 3C~132, \Ha\
in 3C~346 due to a telluric band, and \Hb\ in 3C~458. We complemented our
observations with the data for 3C~270 from \citet{ho97} that were obtained
using a similar extraction aperture and spectral resolution.  Following
\citetalias{buttiglione10alias} we used the Excitation Index (E.I.), defined
as

E.I. = Log [O~III]/\Hb\ - 1/3 (Log [N~II]$/$\Ha + Log [S~II]$/$\Ha + Log
[O~I]$/$\Ha) 

to derive an optical spectroscopic classification of the eight sources where
all six diagnostic lines could be measured. Defining as LEG the sources with
E.I.  $\lesssim$ 0.95, four of them are LEG and four are HEG (see Table
\ref{speclas}).  For 3C~132 we lack the measurement of its [O~I] line;
however, this source is located well within the region of LEG in the
diagnostic diagrams, see Fig.  \ref{diag-lri}, where it is represented by the
filled triangle. Conversely, 3C~346, without a \Ha\ flux estimate, could not
be classified based on the emission line ratios; similarly we cannot derive a
classification from the diagrams comparing lines and radio luminosity since
its location in Fig. \ref{o3re} (the filled triangle in the left panel) is
between the relations defined by LEG and HEG. Instead 3C~458 can be defined as
HEG, although the \Hb\ line cannot be measured, from the lower limit (E.I. $>
1.03$) derived for this source.

Including 3C~410 there are now 19 3CR radio-galaxies with broad lines. In
agreement with our previous findings, also 3C~410 is a HEG from the point of
view of its narrow line ratios. The four newly discovered HEG are of high
total 

\onecolumn
\begin{landscape}

\begin{longtable}{l| c c| c| c | c c| c c c|c}
\caption[Log of the observations.]{Log of the observations.} 
\label{logoss} \\
      \hline \hline
Name &\multicolumn{2}{|c|}{ Coordinates J2000}& z & Obs. Date  & \multicolumn{2}{|c|}{ Low Res.}
&\multicolumn{3}{|c|}{  High Res. }  & Notes \\
\hline
         &  $\alpha$   &  $\delta$      &        &         & n & T$_{exp}$ & HR & n &  T$_{exp}$ &  \\ \hline 
3C~020   & 00 43 09.27 &  +52 03 36.66  & 0.174  & 22Sep08 & 1 & 750 & HRI & 2 & 750 & \\    
3C~063   & 02 20 53.82 & --01 57 54.08  & 0.175  & 23Sep08 & 1 & 750 & HRI & 2 & 750 & \\                                      
3C~132   & 04 56 43.40 &  +22 49 21.62  & 0.214  & 23Oct08 & 2 & 500 & HRI & 2 &1000 & \\    
3C~288   & 13 38 50.00 &  +38 51 10.70  & 0.246  & 18Jan09 & 2 & 500 & HRI & 2 &1000 &  \\   
3C~346   & 16 43 48.69 &  +17 15 48.09  & 0.161  & 01Sep08 & 1 & 750 & HRI & 2 & 750 & i \\  
3C~349   & 16 59 28.84 &  +47 02 56.80  & 0.205  & 30Aug08 & 2 & 500 & HRI & 2 &1000 &  \\   
3C~403.1 & 19 52 30.58 & --01 17 19.68  & 0.055  & 22Sep08 & 1 & 500 & HRR & 2 & 500 &  \\                              
3C~410   & 20 20 06.56 &  +29 42 14.20  & 0.249  & 23Sep08 & 2 & 500 & HRI & 2 &1000 & f,g \\
3C~458	 & 23 12 54.40 &  +05 16 46.00  & 0.289  & 29Lug09 & 2 & 500 & HRI & 2 &1000 &  \\
   \hline
\end{longtable}

Column description: (1) 3C name of the source; (2) and (3) J2000 coordinates
(right ascension and declination); (4) redshift; (5) UT night of observation;
(6) number of low resolution spectra; (7) exposure time for each low
resolution spectrum; (8) high resolution grism used; (9) number of high
resolution spectra; (10) exposure time for each high resolution spectrum.
(11): (f) broad components; (g) no starlight subtraction; (i) telluric
correction.

\bigskip

\begin{longtable}{l c c c c c c c c c c c c}
\caption[Emission line measurements.]{Emission line measurements.} 
\label{bigtable} \\

\hline \hline 
Name    & Redshift  & E(B-V) & L(H$\alpha$) & F(H$\alpha$) & H$\beta$ & [O III]$\lambda$5007 & [O I]$\lambda$6300 & [N II]$\lambda$6584 & [S II]$\lambda$6716 & [S II]$\lambda$6731 & F(H$\alpha$) broad\\
\hline	
3C~020   & 0.175 & 0.407 & 41.37 & -14.55 ( 4)&0.26 (10)& 1.48 ( 2)& 0.15 (17) & 0.91 (3) & 0.33 ( 1)& 0.26 ( 5) &   \\          
3C~063   & 0.173 & 0.027 & 41.54 & -14.38 ( 5)&0.23 ( 4)& 1.22 ( 1)& 0.23 ( 3) & 0.29 (2) & 0.18 ( 4)& 0.13 ( 5) &   \\          
3C~132	 & 0.201 & 0.482 & 41.37 & -14.75 ( 1)&0.90 ( 5)& 1.24 ( 2)& $<$0.20   & 1.78 (1) & 0.67 ( 2)& 0.73 ( 2) &   \\          
3C~288	 & 0.245 & 0.007 & 40.86 & -15.40 ( 6)&0.58 (17)& 0.62 (16)& 0.40 (13) & 1.87 (3) & 0.85 ( 7)& 0.60 ( 9) &   \\          
3C~346 &0.163&0.067& 41.24$^a$&-14.62$^a$ ( 1)&0.20$^a$ (10)&1.00$^a$ ( 2)&0.14$^a$ (14)&0.84$^a$ ( 4)&0.13$^a$ (11)& 0.13$^a$ (11) &  \\
3C~349   & 0.206 & 0.031 & 41.50 & -14.58 ( 2)&0.40 ( 6)& 1.54 ( 2)& 0.25 (13) & 0.90 (2) & 0.34 ( 5)& 0.28 ( 8) &   \\          
3C~403.1 & 0.056 & 0.234 & 39.99 & -14.87 (15)&0.53 (13)& 0.75 ( 7)& 0.29 (17) & 0.78 (6) & 0.90 ( 3)& 0.72 (11) &   \\          
3C~410   & 0.249 & 0.050 & 41.86 & -14.40 ( 8)&0.12 (13)& 1.46 ( 1)& 0.18 ( 9) & 1.04 (1) & 0.37 ( 3)& 0.30 ( 3) & -13.47 (1)  \\
3C~458.0 & 0.289 & 0.082 & 41.58 & -14.84 ( 4)&$<$0.45  & 2.85 ( 3)& 0.33 (14) & 0.91 ( 3)& 0.36 ( 4)& 0.31 ( 8) &   \\
\hline
3C~270   & 0.007 & 0.018 & 39.26 & -13.80     & 0.20    & 0.51     & 0.49     & 2.60      & 0.72     & 0.57      & \\
\hline
\hline
\end{longtable}
Column description: (1) source name; (2) redshift; (3) Galactic absorption;
(4) logarithm of the luminosity of the H$\alpha$ narrow line, in erg s$^{-1}$;
(5) logarithm of the \Ha\ flux in erg cm$^{-2}$ s$^{-1}$; (6 through 11)
de-reddened flux ratios of the key diagnostic lines with respect to H$\alpha$.
The values in parentheses report the errors (in percentage) of each line;
(12) logarithm of the flux of the \Ha\ broad component, when visible. Notes:
(a) for 3C~346 no measurement of \Ha\ is possible and we give instead the [O
III] luminosity, referring the flux ratios to this line. Data for 3C~270 are taken from
\citet{ho97} and corrected for Galactic reddening.
\end{landscape}
\twocolumn

\noindent
radio-power, Log $L_{178} > 32.8$ [$\ergsHz$], and with a FR~II
morphology, a result that applies to all object of this spectroscopic
sub-class.

The number of LEG with high radio luminosity, Log $L_{178} > 34$ [$\ergsHz$],
and a FR~II morphology is significantly increased by the new observations
(with the additions of 3C~132, 3C~288, and 3C~349), from six to nine objects,
confirming the relevance of this subclass.

In \citetalias{buttiglione10alias} we derived the best fit correlation for the
link between line luminosity with radio power, considering separately the
sub-populations of HEG and LEG. Including the new sources we find:

Log $L_{\rm[O~III]}$ = 1.10 $~$Log $L_{178}$ + 4.54 (for HEG) and 

Log $L_{\rm[O~III]}$ = 0.98 $~$Log $L_{178}$ + 7.86 (for LEG).

Considering instead the \Ha\ line we have

Log $L_{\rm {H\alpha}}$ = 1.01 $~$Log $L_{178}$ + $\,\,\,$7.35 (for HEG) and

Log $L_{\rm {H\alpha}}$ = 0.82 $~$Log $L_{178}$ + 13.31 (for LEG).

The slopes of the correlations are only marginally reduced with respect to the
values reported in Paper I, by 0.05 and 0.01 for HEG and LEG respectively, for
both lines. The errors in the slopes are also marginally reduced to 0.10
(0.09) for the relation between $L_{\rm[O~III]}$ and Log $L_{178}$ for HEG
(LEG).

\begin{table*}
  \begin{center}
    \caption{Multiwavelength data and spectroscopic classification}
    \label{speclas}
    \begin{tabular}{l| c| c c| c c| c| c c c}
      \hline \hline
Name & redshift & \multicolumn{2}{|c|}{ Emission lines}& \multicolumn{2}{|c|}{
  Radio emission}  & Host magnitude & \multicolumn{3}{|c}{Classification}  \\
\hline
     & &  H$\alpha$  & [O~III] & L$_{178}$ &  P$_{core}$ & M$_{H}$ &  FR & spec & Method\\ 
\hline 
3C~020   &  0.174 & 41.37&  41.54 & 34.55 &30.44 &  -24.64$^*$& 2 & HEG & E.I. \\
3C~063   &  0.175 & 41.54&  41.63 & 34.21 &31.12 &    --      &   & HEG & E.I. \\
3C~132   &  0.214 & 41.37&  41.46 & 34.25 &31.58 &  -26.00    & 2 & LEG & D.D. \\
3C~270	 &  0.007 & 39.26&  38.96 & 31.79 &29.57 &  -25.01    & 1 & LEG & E.I. \\
3C~288   &  0.246 & 40.86&  40.65 & 34.53 &31.73 &  -26.10$^*$& 2 & LEG & E.I. \\
3C~346   &  0.161 &   -- &  41.24 & 33.88 &32.18 &  -25.84    & 2 &  -- &      \\
3C~349   &  0.205 & 41.50&  41.69 & 34.20 &31.35 &  -24.82$^*$& 2 & LEG & E.I. \\
3C~403.1 &  0.055 & 39.99&  39.86 & 32.98 & --   &  -24.36    &   & LEG & E.I. \\
3C~410   &  0.248 & 41.86&  42.02 & 34.80 &33.43 &    --      & 2 & BLO & E.I. \\
3C~458   &  0.289 & 41.58&  42.03 & 34.58 &30.88 &    --      & 2 & HEG & E.I. \\
\hline                                                          
    \end{tabular}                                               
  \end{center}                                                  
  Column description: (1) 3CR name; (2) redshift from \citet{spinrad85}; (3) and (4) logarithm
  of H$\alpha$ and [O~III]$\lambda$5007 luminosities [erg s$^{-1}$]; (5) radio luminosity at
  178 MHz [erg s$^{-1}$ Hz$^{-1}$] from \citet{spinrad85}; (6) radio core power at 5 GHz [erg
  s$^{-1}$ Hz$^{-1}$] from \citet{baldi09}; (7) host
  H magnitude from 2MASS \citep{skrutskie06} 
(or from HST \citep{donzelli07} for the objects marked with a $^*$);
  (8): morphological FR type; (9) spectroscopic classification into High
  Excitation Galaxy (HEG); 
  Low Excitation Galaxy (LEG); Broad Line Object (BLO); (--) unclassified.  Column (10)
  classification method: E.I. - excitation index; D.D. - diagnostic diagrams. 
\end{table*}

\section{Summary}
\label{summary}

We presented optical spectroscopic data of nine 3CR radio sources, needed to
complete our survey of this catalogue with redshift $<$ 0.3, and measured
emission lines luminosities and ratios. These data enabled us to derive an
optical spectroscopic classification for all but one galaxy.  The
relationships between spectroscopic and radio properties found from our
previous works, are confirmed by the analysis of the now complete sample.
\begin{acknowledgements}
SB and ACe acknowledge the Italian MIUR for financial
support. ACa acknowledges COFIN-INAF-2006 grant financial
support. This research has made use of the NASA/IPAC
Extragalactic Database (NED) which is operated by the Jet Propulsion
Laboratory. California institute of Technology, under contract with the
National Aeronautics and Space Administration.  This research has made use of
NASA's Astrophysics Data System (ADS).  
\end{acknowledgements} 
\bibliographystyle{aa}

\clearpage

\appendix
\section{Additional material}
In the following two tables we provide the data for the whole 3CR sub-sample
of radio galaxies with $z<0.3$ combining the results presented here with those
of \citetalias{buttiglione09alias} and \citetalias{buttiglione10alias}.  This
is additional material with respect to the article published in Astronomy \&
Astrophysics.

\input tab1.tex

\input tab2.tex

\end{document}

%% file: tab1.tex
\onecolumn
\begin{landscape}
\begin{longtable}{l c c c c c c c c c c c c}

\caption[Emission line measurements.]{Emission line measurements.} 
\label{fulltable1} \\

\hline \hline 
Name     & Redshift   & E(B-V) & L(H$\alpha$) & F(H$\alpha$) & H$\beta$  & [O III]$\lambda$5007 & [O I]$\lambda$6364 & [N II]$\lambda$6584 & [S II]$\lambda$6716 & [S II]$\lambda$6731 & F(H$\alpha$) broad\\
\hline	
\endfirsthead

\multicolumn{3}{c}{{\tablename} \thetable{} -- Continued} \\[0.5ex]
\hline \hline 
Name     & Redshift   & E(B-V) & L(H$\alpha$) & F(H$\alpha$) & H$\beta$  & [O III]$\lambda$5007 & [O I]$\lambda$6364 & [N II]$\lambda$6584 & [S II]$\lambda$6716 & [S II]$\lambda$6731 & F(H$\alpha$) broad\\
\hline
\endhead

\hline
  \multicolumn{10}{c}{{Continued on Next Page}} \\
\endfoot

  \\[-1.8ex] 
\endlastfoot


3C~015.0 & 0.073  &  0.022 &  40.40 &    -14.70 ( 2) &   0.32 (20) &   1.58 ( 4) &   0.29 ( 8) &   2.06 ( 1) &   0.32 ( 8) &   0.41 ( 8)  & \\
3C~017.0 & 0.220  &  0.023 &  41.88 &    -14.27 ( 1) &   0.24 ( 8) &   1.28 ( 1) &   0.44 ( 1) &   0.70 ( 1) &   0.23 ( 1) &   0.21 ( 2)  & -13.87 \\
3C~018.0 & 0.188  &  0.158 &  41.93 &    -14.06 ( 1) &   0.33 ( 1) &   4.17 ( 1) &   0.43 ( 3) &   1.13 ( 1) &   0.38 ( 2) &   0.42 ( 1)  & -13.03 \\
3C~020.0  & 0.175 & 0.407 & 41.37 & -14.55 ( 4)&0.26 (10)& 1.48 ( 2)& 0.15 (17) & 0.91 (3) & 0.33 ( 1)& 0.26 ( 5) &   \\          
3C~028.0 & 0.195  &  0.058 &  41.51 &    -14.52 ( 2) &   0.53 (15) &   0.28 (20) &   0.17 (10) &   0.92 ( 3) &   0.55 (10) &          --  & \\
3C~029.0 & 0.045  &  0.036 &  40.06 &    -14.60 ( 6) &   0.24 (25) &   1.07 ( 2) &   0.19 (16) &   1.85 ( 1) &   0.50 ( 6) &   0.52 ( 4)  & \\
3C~031.0 & 0.017  &  0.001 &  39.83 &    -13.96 ( 1) &   0.15 ( 8) &   0.43 ( 2) &   0.14 (12) &   0.99 ( 1) &   0.37 ( 1) &   0.32 ( 1)  & \\
3C~033.0 & 0.060  &  0.028 &  41.63 &    -13.29 ( 1) &   0.31 ( 1) &   3.55 ( 1) &   0.25 ( 1) &   0.63 ( 1) &   0.39 ( 1) &   0.33 ( 1)  & \\
3C~033.1 & 0.181  &  0.633 &  41.85 &    -14.11 ( 1) &   0.22 ( 5) &   2.80 ( 1) &   0.25 ( 1) &   0.57 ( 1) &   0.27 ( 1) &   0.21 ( 1)  & -13.26  \\
3C~035.0 & 0.067  &  0.141 &  40.22 &    -14.81 ( 1) & $<$  1.27   &   0.62 (23) &   0.46 ( 5) &   0.77 ( 2) &   0.62 (15) &          --  & \\
3C~040.0 & 0.019  &  0.041 &  39.08 &    -14.79 (11) &   0.32 (32) &   1.38 ( 6) &   0.24 (28) &   2.32 ( 1) &   0.81 (10) & $<$  0.52    & \\
3C~052.0 & 0.285  & 0.232  &$<$ 40.64 & $<$ -15.76   &     --      &         --  &         --  &        --   &        --   &          --  & \\
3C~061.1 & 0.184  &  0.176 &  42.05 &    -13.92 ( 1) &   0.25 ( 4) &   2.63 ( 1) &   0.08 (10) &   0.31 ( 1) &   0.16 ( 1) &   0.15 ( 3)  & \\
3C~063.0 & 0.173 & 0.027 & 41.54 & -14.38 ( 5)&0.23 ( 4)& 1.22 ( 1)& 0.23 ( 3) & 0.29 (2) & 0.18 ( 4)& 0.13 ( 5) &   \\          
3C~066B  & 0.022  &  0.080 &  40.11 &    -13.90 ( 4) &   0.22 ( 7) &   0.87 ( 1) &   0.26 (16) &   2.45 ( 1) &   0.56 ( 9) &       --     & \\
3C~075N  & 0.023  &  0.180 &  39.58 &    -14.50 ( 1) & $<$  2.20   & $<$  2.20   &   0.42 ( 2) &   2.48( 1)  & $<$  0.69   &   0.37 ( 1)  & \\
3C~076.1 & 0.033  &  0.138 &  39.89 &    -14.50 ( 2) & $<$  0.85   & $<$  0.92   & $<$  0.18   &   1.57 ( 1) &   0.33 ( 4) &   0.54 ( 1)  & \\
3C~078.0 & 0.029  &  0.173 &  39.73 &    -14.53 ( 3) &   0.18 (30) &   0.48 (16) &   0.18 (13) &   1.88 ( 2) &       --    &       --     & \\
3C~079.0 & 0.256  &  0.127 &  42.39 &    -13.91 ( 1) &   0.29 ( 3) &   2.97 ( 1) &   0.05 ( 2) &   0.32 ( 1) &   0.12 ( 3) &   0.11 ( 7)  & \\
3C~083.1 & 0.027  &  0.164 &  39.40 &    -14.83 (14) & $<$  0.72   & $<$  1.25   &       --    &   1.35 ( 3) &       --    &       --     & \\
3C~084.0 & 0.018  &  0.163 &  41.28 &    -12.55 ( 1) &   0.42 ( 1) &   2.09 ( 1) &   0.64 ( 1) &   1.12 ( 1) &   0.54 ( 1) &   0.51 ( 1)  & \\
3C~088.0 & 0.030  &  0.126 &  39.98 &    -14.33 ( 1) &   0.29 (11) &   1.44 ( 2) &   0.50 ( 3) &   2.39 ( 1) &   0.97 ( 1) &   0.79 ( 3)  & \\
3C~089.0 & 0.139  & 0.134  &  40.28 &    -15.42 (11) & $<$  1.86   & $<$  1.69   & $<$  1.26   &   1.43 ( 7) &       --    &       --     & \\
3C~093.1 & 0.243  &  0.389 &  42.35 &    -13.89 ( 1) &   0.28 ( 4) &   2.08 ( 1) &   0.28 ( 3) &   1.36 ( 1) &   0.54 ( 1) &   0.55 ( 1)  & \\
3C~098.0 & 0.030  &  0.229 &  40.52 &    -13.79 ( 1) &   0.25 ( 3) &   3.01 ( 1) &   0.15 ( 3) &   0.76 ( 1) &   0.34 (10) &   0.23 ( 7)  & \\
3C~105.0 & 0.089  &  0.480 &  40.89 &    -14.39 ( 3) &   0.26 (28) &   3.59 ( 1) &   0.38 ( 5) &   1.59 ( 1) &   0.55 ( 4) &   0.55 ( 1)  & \\
3C~111.0 & 0.049  &  1.647 &42.44$^a$& -12.28$^a$ (1)   &   --     & 1.00$^a$( 1)&   0.04$^a$ (10)  &  -- &  0.03$^a$ ( 7) & 0.03$^a$ ( 9 )& -11.64 \\ 
3C~123.0 & 0.218  &  0.981 &  41.96 &    -14.18 ( 1) &   0.61 (32) &   1.09 (18) &   0.25 (32) &   2.34 ( 1) &   0.48 ( 1) &   0.35 ( 1)  & \\
3C~129.0 & 0.022  &  1.058 &  39.81 &    -14.20 ( 1) & $<$  0.99   & $<$  1.10   &   0.61 ( 1) &   1.51 ( 1) &   0.50 ( 1) &   0.50 (20)  & \\
3C~129.1 & 0.022  &  1.131 & $<$ 39.83 & $<$ -14.21  &      --     &         --  &         --  &        --   &        --   &          --  & \\
3C~130.0 & 0.032  &  1.309 & $<$ 40.17 & $<$ -14.19  &      --     &         --  &         --  &        --   &        --   &          --  & \\
3C~132.0 & 0.201 & 0.482 & 41.37 & -14.75 ( 1)&0.90 ( 5)& 1.24 ( 2)& $<$0.20   & 1.78 (1) & 0.67 ( 2)& 0.73 ( 2) &   \\          
3C~133.0 & 0.278  &  0.949 &  42.41 &    -13.97 ( 1) &   0.32 ( 1) &   2.26 ( 1) &   0.25 ( 9) &   0.70 ( 1) &   0.17 ( 1) &   0.14 ( 1)  & \\
3C~135.0 & 0.125  &  0.115 &  41.52 &    -14.09 ( 1) &   0.33 ( 3) &   3.40 ( 1) &   0.18 ( 2) &   0.80 ( 1) &   0.30 ( 1) &   0.30 ( 3)  & \\
3C~136.1 & 0.064  &  0.762 &  41.41 &    -13.57 ( 1) &   0.14 ( 1) &   1.08 ( 1) &   0.05 ( 1) &   0.59 ( 1) &   0.20 ( 1) &   0.20 ( 3)  & \\
3C~153.0 & 0.277  &  0.162 &  41.60 &    -14.77 ( 3) &   0.23 (15) &   1.07 ( 3) &   0.36 ( 7) &   1.21 ( 3) &   0.54 ( 8) &   0.33 (13)  & \\
3C~165.0 & 0.296  &  0.174 &  41.44 &    -15.00 (15) &   0.44 (15) &   1.68 ( 4) &   0.31 (32) &   1.14 (27) &   0.32 (27) &   0.33 (27)  & \\
3C~166.0 & 0.245  &  0.211 &  41.51 &    -14.75 ( 3) &   0.42 ( 6) &   1.40 ( 2) &   0.38 ( 7) &   0.73 ( 4) &   0.41 (16) &   0.34 (24)  & \\
3C~171.0 & 0.238  &  0.054 &  42.45 &    -13.78 ( 1) &   0.36 ( 1) &   2.73 ( 1) &   0.24 ( 2) &   0.57 ( 1) &   0.38 ( 6) &   0.29 ( 2)  & \\
3C~173.1 & 0.292  &  0.044 &  41.05 &    -15.38 ( 7) &  $<$   0.23 &   0.63 (12) &   0.24 (28) &   2.04 ( 3) &   0.24 (29) &   0.31 (24)  & \\
3C~180.0 & 0.220  &  0.098 &  41.79 &    -14.36 ( 1) &   0.25 ( 4) &   3.53 ( 1) &   0.10 (28) &   0.69 ( 1) &   0.28 ( 3) &   0.19 ( 5)  & \\
3C~184.1 & 0.118  &  0.032 &  41.79 &    -13.77 ( 3) &   0.29 ( 4) &   2.71 ( 1) &   0.08 ( 7) &   0.27 ( 8) &   0.11 ( 1) &   0.09 ( 3)  & -13.99 \\
3C~192.0 & 0.060  &  0.054 &  40.95 &    -13.97 ( 1) &   0.30 ( 1) &   2.48 ( 1) &   0.11 ( 3) &   0.71 ( 1) &   0.35 ( 1) &   0.26 ( 1)  & \\
3C~196.1 & 0.198  &  0.065 &  41.56 &    -14.48 ( 2) &   0.22 (10) &   0.91 ( 3) &   0.20 (15) &   1.19 ( 1) &   0.52 ( 1) &   0.49 ( 1)  & \\
3C~197.1 & 0.128  &  0.041 &  40.69 &    -14.93 ( 3) &   0.37 (11) &   1.69 ( 2) &   0.33 ( 8) &   0.76 ( 4) &   0.29 (11) &   0.31 ( 9)  & -13.95 \\
3C~198.0 & 0.082  &  0.026 &  41.31 &    -13.89 ( 1) &   0.27 ( 1) &   0.46 ( 1) &   0.05 ( 9) &   0.28 ( 1) &       --    &       --     & \\
3C~213.1 & 0.194  &  0.028 &  41.01 &    -15.02 ( 3) &   0.21 (14) &   1.13 ( 3) &   0.46 ( 6) &   1.41 ( 2) &   0.38 (10) &   0.28 (12)  & \\
3C~219.0 & 0.175  &  0.018 &  41.55 &    -14.38 ( 2) &   0.25 (10) &   1.67 ( 1) &   0.44 ( 3) &   0.90 ( 1) &   0.24 ( 6) &   0.24 ( 7)  & -13.87 \\
3C~223.0 & 0.137  &  0.012 &  41.68 &    -14.01 ( 1) &   0.23 ( 5) &   3.09 ( 1) &   0.19 ( 3) &   0.63 ( 1) &   0.25 ( 3) &   0.21 ( 3)  & \\
3C~223.1 & 0.107  &  0.017 &  41.16 &    -14.30 ( 1) &   0.28 ( 7) &   2.63 ( 1) &   0.06 (15) &   0.81 ( 1) &   0.22 ( 4) &   0.19 ( 5)  & \\
3C~227.0 & 0.086  &  0.026 &  41.08 &    -14.17 ( 2) &   0.44 ( 1) &   4.73 ( 1) &   0.22 (13) &   0.19 ( 4) &   0.16 ( 5) &   0.16 (11)  & -12.52 \\
3C~234.0 & 0.185  &  0.019 &  42.64 &    -13.33 ( 1) &   0.25 ( 2) &   2.96 ( 1) &   0.04 ( 8) &   0.28 ( 1) &   0.08 ( 3) &   0.07 ( 1)  & -13.29 \\
3C~236.0 & 0.099  &  0.011 &  41.13 &    -14.25 ( 1) &   0.22 ( 4) &   0.57 ( 2) &   0.30 ( 3) &   0.69 ( 1) &   0.49 ( 2) &   0.35 ( 3)  & \\
3C~258.0 & 0.165  &  0.020 &  40.96 &    -14.90 (23) &   0.11 (24) &   0.17 (15) & $<$  0.47   & $<$  2.14   &       --    &       --     & \\
3C~264.0 & 0.022  &  0.023 &  39.68 &    -14.35 ( 1) &   0.27 ( 7) &   0.33 (10) &   0.22 ( 9) &   1.45 ( 1) &   0.33 ( 8) &   0.33 (22)  & \\
3C~270.0 & 0.007  & 0.018  &  39.26 &    -13.80 (--) &   0.20 (--) &   0.51 (--) &   0.49 (--) &   2.60 (--) &   0.72 (--) &   0.57 (--)  & \\
3C~272.1 & 0.004  &  0.040 &  38.92 &    -13.57 ( 1) &   0.10 ( 3) &   0.19 ( 4) &   0.23 ( 6) &   1.28 ( 1) &   0.52 ( 1) &   0.34 ( 1)  & \\
3C~273.0 & 0.158  &  0.021 &   --    &    --           &   --         &    --        &        --    &  --          &   --         &   --  & -11.51 \\ 
3C~274.0 & 0.004  &  0.022 &  39.50 &    -13.11 ( 1) &   0.17 ( 1) &   0.31 ( 1) &   0.36 ( 1) &   2.32 ( 1) &   0.68 ( 1) &   0.77 ( 1)  & \\
3C~277.3 & 0.086  &  0.012 &  40.83 &    -14.43 ( 1) &   0.19 ( 9) &   1.29 ( 1) &   0.29 ( 5) &   0.79 ( 3) &   0.32 ( 1) &   0.28 ( 5)  & \\
3C~284.0 & 0.239  &  0.016 &  41.41 &    -14.82 ( 7) &   0.20 (12) &   1.52 ( 1) & $<$  0.24   &   0.61 (11) & $<$  0.45   & $<$  0.41    & \\
3C~285.0 & 0.079  &  0.017 &  40.66 &    -14.52 ( 1) &   0.19 ( 6) &   0.78 ( 1) &   0.10 (10) &   0.54 ( 2) &   0.27 ( 4) &   0.19 ( 6)  & \\
3C~287.1 & 0.216  &  0.025 &  41.50 &    -14.62 ( 3) &   0.27 ( 9) &   1.71 ( 1) &   0.48 ( 4) &   0.68 ( 4) &   0.29 ( 7) &   0.29 ( 7)  & -13.85 \\
3C~288.0 & 0.245 & 0.007 & 40.86 & -15.40 ( 6)&0.58 (17)& 0.62 (16)& 0.40 (13) & 1.87 (3) & 0.85 ( 7)& 0.60 ( 9) &   \\          
3C~293.0 & 0.045  &  0.017 &  40.18 &    -14.49 ( 3) &   0.19 (17) &   0.42 (10) &   0.26 ( 3) &   0.88 ( 1) &   0.66 ( 2) &   0.66 (10)  & \\
3C~296.0 & 0.025  &  0.025 &  39.87 &    -14.28 ( 1) &   0.30 (12) &   0.81 ( 2) &   0.22 (23) &   1.84 ( 1) &   0.43 ( 2) &   0.38 (10)  & \\
3C~300.0 & 0.272  &  0.035 &  41.78 &    -14.58 ( 2) &   0.25 ( 9) &   1.71 ( 1) &   0.15 (13) &   0.48 ( 4) &   0.33 ( 4) &   0.23 ( 4)  & \\
3C~303.0 & 0.141  &  0.019 &  41.33 &    -14.39 ( 1) &   0.35 ( 4) &   2.55 ( 1) &   0.41 ( 2) &   1.06 ( 1) &   0.46 ( 2) &   0.39 ( 3)  & -13.46 \\
3C~303.1 & 0.269  &  0.036 &  42.10 &    -14.24 ( 1) &   0.26 ( 2) &   2.07 ( 1) &   0.24 ( 3) &   1.03 ( 1) &   0.38 ( 2) &   0.48 ( 1)  & \\
3C~305.0 & 0.042  &  0.026 &  40.92 &    -13.68 ( 1) &   0.12 ( 4) &   1.30 ( 1) &   0.17 ( 7) &   1.77 ( 1) &   0.48 ( 1) &   0.40 ( 1)  & \\
3C~310.0 & 0.054  &  0.042 &  40.32 &    -14.50 ( 1) &   0.23 ( 7) &   0.54 ( 3) &   0.30 ( 4) &   1.74 ( 1) &   0.84 ( 1) &   0.74 ( 2)  & \\
3C~314.1 & 0.120  &  0.020 &  40.31 &    -15.25 ( 4) &   0.37 (16) &   0.24 (24) &   0.18 (23) &   0.53 ( 7) &   0.52 ( 8) &   0.33 (13)  & \\
3C~315.0 & 0.108  &  0.062 &  41.15 &    -14.32 ( 1) &   0.20 ( 4) &   0.53 ( 1) &   0.25 ( 2) &   0.72 ( 1) &   0.51 ( 1) &   0.37 ( 2)  & \\
3C~317.0 & 0.034  &  0.037 &  40.35 &    -14.08 ( 1) &   0.30 ( 5) &   1.00 ( 2) &   0.25 ( 5) &   1.92 ( 1) &   0.58 ( 1) &   0.49 ( 1)  & \\
3C~318.1 & 0.044  &  0.035 &  39.95 &    -14.70 ( 4) &  $<$ 0.58   &   0.26 (32) &   0.21 (19) &   1.02 ( 3) &   0.29 ( 9) &   0.16 (13)  & \\
3C~319.0 & 0.189  & 0.012  &  41.16 &    -14.84 ( 7) & $<$  0.13   & $<$  0.10   & $<$  0.27   &   0.30 ( 6) &   0.15 (25) & $<$  0.14    & \\
3C~321.0 & 0.097  &  0.044 &  40.50 &    -14.87 ( 2) &   0.30 (11) &   2.58 ( 1) &   0.06 (28) &   0.50 ( 3) &   0.18 ( 8) &   0.15 (11)  & \\
3C~323.1 & 0.264  &  0.042 &  42.21 &    -14.12 ( 1) &   0.26 (11) &   3.93 ( 1) &   0.13 (20) &   0.29 (32) &   0.10 (11) &   0.08 (21)  & -12.37 \\
3C~326.0 & 0.090  &  0.053 &  40.28 &    -15.02 ( 5) & $<$  0.23   &   1.31 ( 4) &   0.37 (13) &   1.93 ( 3) &   0.41 (12) &   0.50 (11)  & \\
3C~327.0 & 0.104  &  0.089 &  41.73 &    -13.70 ( 1) &   0.27 ( 4) &   3.20 ( 1) &   0.14 ( 1) &   0.73 ( 1) &   0.30 ( 1) &   0.24 ( 1)  & \\
3C~332.0 & 0.151  &  0.024 &  41.31 &    -14.47 ( 2) &   0.28 ( 8) &   3.14 ( 1) &   0.21 ( 9) &   0.97 ( 1) &   0.37 ( 6) &   0.36 ( 7)  & -12.77 \\
3C~338.0 & 0.032  &  0.012 &  40.25 &    -14.11 ( 1) &   0.18 (19) &   0.21 ( 6) &   0.18 ( 2) &   1.63 ( 1) &   0.41 ( 1) &   0.33 ( 1)  & \\
3C~346.0 & 0.163 & 0.067 & 41.24$^a$&-14.62$^a$ ( 1)&0.20$^a$ (10)&1.00$^a$ ( 2)&0.14$^a$ (14)&0.84$^a$ ( 4)&0.13$^a$ (11)& 0.13$^a$ (11) &  \\
3C~348.0 & 0.154  &  0.094 &  41.29 &    -14.51 ( 1) &   0.25 ( 5) &   0.13 ( 9) &   0.27 ( 5) &   1.27 ( 1) &       --    &       --     & \\
3C~349.0 & 0.206 & 0.031 & 41.50 & -14.58 ( 2)&0.40 ( 6)& 1.54 ( 2)& 0.25 (13) & 0.90 (2) & 0.34 ( 5)& 0.28 ( 8) &   \\          
3C~353.0 & 0.030  &  0.439 &  40.42 &    -13.90 ( 1) &   0.20 (17) &   0.53 ( 7) &   0.30 ( 2) &   1.09 ( 1) &   0.56 ( 1) &   0.43 ( 1)  & \\
3C~357.0 & 0.166  &  0.045 &  40.92 &    -14.96 ( 3) &   0.23 (16) &   1.08 ( 4) &   0.46 ( 7) &   1.37 ( 4) &       --    &       --     & \\
3C~371.0 & 0.050  &  0.036 &  40.94 &    -13.82 ( 1) &   0.29 (10) &   1.01 ( 4) &   0.51 ( 2) &   1.14 ( 1) &   0.32 ( 3) &   0.29 ( 1)  & \\
3C~379.1 & 0.256  &  0.062 &  41.41 &    -14.89 ( 6) &   0.31 ( 6) &   2.80 ( 1) &   0.21 (19) &   1.54 ( 5) &   0.30 (13) &   0.28 (14)  & \\
3C~381.0 & 0.161  &  0.053 &  41.79 &    -14.05 ( 1) &   0.31 ( 2) &   3.83 ( 1) &   0.11 ( 9) &   0.53 ( 1) &       --    &       --     & \\
3C~382.0 & 0.058  &  0.070 &  41.39 &    -13.51 ( 1) &   0.31 ( 1) &   2.45 ( 1) &   0.20 ( 6) &   1.49 ( 1) &   0.12 ( 8) &   0.13 ( 9)  & -11.61 \\
3C~386.0 & 0.017  &  0.335 &  40.17 &    -13.63 ( 1) & $<$  1.08   & $<$  1.08   &   0.10 (17) &   0.57 ( 1) &   0.11 ( 7) &   0.08 (10)  & \\
3C~388.0 & 0.091  &  0.080 &  40.83 &    -14.47 ( 2) &   0.23 (14) &   0.74 ( 3) &   0.26 (13) &   2.33 ( 1) &   0.41 ( 1) &   0.37 ( 1)  & \\
3C~390.3 & 0.056  &  0.071 &  41.57 &    -13.29 ( 1) &   0.32 ( 1) &   3.24 ( 1) &   0.27 ( 1) &   0.47 ( 1) &   0.10 ( 1) &   0.10 ( 1)  & -11.60 \\
3C~401.0 & 0.201  &  0.059 &  41.01 &    -15.05 ( 3) &   0.30 (15) &   1.10 ( 5) &   0.24 (16) &   1.77 ( 2) &   0.46 ( 9) &   0.34 (14)  & \\
3C~402.0 & 0.024  &  0.121 &  39.08 &    -15.03 ( 3) & $<$  1.95   & $<$  2.19   &   0.37 (13) &   2.97 ( 1) & $<$  0.44   & $<$  0.97    & \\
3C~403.0 & 0.059  &  0.187 &  41.20 &    -13.71 ( 1) &   0.25 ( 3) &   3.54 ( 1) &   0.13 ( 3) &   0.84 ( 1) &   0.25 ( 1) &   0.24 ( 2)  & \\
3C~403.1 & 0.056 & 0.234 & 39.99 & -14.87 (15)&0.53 (13)& 0.75 ( 7)& 0.29 (17) & 0.78 (6) & 0.90 ( 3)& 0.72 (11) &   \\          
3C~410.0 & 0.249 & 0.050 & 41.86 & -14.40 ( 8)&0.12 (13)& 1.46 ( 1)& 0.18 ( 9) & 1.04 (1) & 0.37 ( 3)& 0.30 ( 3) & -13.47 (1)  \\
3C~424.0 & 0.127  &  0.096 &  41.07 &    -14.55 ( 1) &   0.24 ( 6) &   0.54 ( 3) &   0.27 ( 5) &   0.79 ( 1) &   0.44 ( 3) &   0.40 (12)  & \\
3C~430.0 & 0.054  &  0.630 &  40.12 &    -14.72 ( 3) & $<$ 1.09    &   1.61 ( 9) &   0.33 (20) &   1.43 ( 1) &   0.37 ( 8) &   0.40 ( 4)  & \\
3C~433.0 & 0.102  &  0.145 &  41.40 &    -14.01 ( 1) &   0.19 ( 4) &   1.88 ( 1) &   0.22 ( 3) &   1.09 ( 1) &   0.46 ( 2) &   0.30 ( 1)  & \\
3C~436.0 & 0.214  &  0.089 &  41.07 &    -15.06 (10) &   0.20 (22) &   3.09 ( 1) &       --    &   2.08 ( 9) &   0.63 (19) &   0.49 (18)  & \\
3C~438.0 & 0.290  &  0.358 &  41.55 &    -14.87 ( 1) & $<$  0.54   & $<$  0.82   & $<$  0.67   &   1.61 ( 1) &  $<$  0.66  &   0.46 (28)  & \\
3C~442.0 & 0.026  &  0.065 &  39.78 &    -14.40 ( 1) &   0.08 (24) &   0.27 ( 7) &   0.19 ( 8) &   1.84 ( 1) &   0.35 ( 5) &   0.45 ( 1)  & \\
3C~445.0 & 0.056  &  0.083 &42.50$^a$&   -12.37$^a$ ( 1)  & -- & 1.00$^a$ ( 1) & 0.04$^a$ ( 6)   &  --  &   0.02$^a$ (10) &  0.02$^a$ (10) & -12.03\\ 
3C~449.0 & 0.017  &  0.167 &  39.71 &    -14.09 ( 1) &   0.10 (23) &   0.30 (24) &   0.13 ( 9) &   1.38 ( 1) &   0.29 ( 7) &   0.22 ( 1)  & \\
3C~452.0 & 0.081  &  0.137 &  41.16 &    -14.05 ( 1) &   0.23 ( 5) &   1.53 ( 1) &   0.27 ( 2) &   1.08 ( 1) &   0.36 ( 1) &   0.29 ( 1)  & \\
3C~456.0 & 0.233  &  0.038 &  42.48 &    -13.72 ( 1) &   0.32 ( 1) &   2.15 ( 1) &   0.15 ( 2) &   0.78 ( 1) &   0.22 ( 1) &   0.24 ( 1)  & \\
3C~458.0 & 0.289  &  0.082 &  41.58 & -14.84 ( 4)&$<$0.45  & 2.85 ( 3)& 0.33 (14) & 0.91 ( 3)& 0.36 ( 4)& 0.31 ( 8) &   \\
3C~459.0 & 0.220  &  0.066 &  42.13 &    -14.02 ( 1) &   0.18 ( 6) &   0.82 ( 1) &   0.14 ( 1) &   1.36 ( 1) &   0.45 ( 1) &   0.30 ( 1)  & -13.83 \\
3C~459.0 & 0.220  &  0.066 &  42.17 &    -13.97 ( 1) &   0.16 ( 6) &   0.73 ( 1) &   0.12 ( 1) &   1.77 ( 1) &   0.35 ( 1) &   0.33 ( 1)  & \\
3C~460.0 & 0.269  &  0.092 &  42.09 &    -14.25 ( 3) &   0.25 ( 3) &   0.49 ( 7) &   0.39 ( 4) &   1.23 ( 2) &   0.60 ( 1) &   0.54 ( 4)  & \\
3C~465.0 & 0.030  &  0.069 &  40.15 &    -14.17 ( 1) &   0.17 (27) &   0.46 ( 6) &   0.26 ( 9) &   2.77 ( 1) &   0.47 ( 1) &   0.32 ( 1)  & \\

\hline
\hline
\end{longtable}
Column description: (1) source name; (2) redshift; (3) Galactic absorption;
(4) logarithm of the luminosity of the H$\alpha$ narrow line, in erg s$^{-1}$;
(5) logarithm of the \Ha\ flux in erg cm$^{-2}$ s$^{-1}$; (6 through 11)
de-reddened flux ratios of the key diagnostic lines with respect to H$\alpha$.
The values in parentheses report the errors (in percentage) of each line.
Missing values marked with `--' correspond to lines outside the coverage of
the spectra or severely affected by telluric bands. When no lines are visible
we only give the upper limit for \Ha; (12) logarithm of the flux of the \Ha\
broad component, when visible. \\ Notes: (a) for 3C~111, 3C~346, and 3C~445,
no narrow \Ha\ measurement is possible and we give instead the [O III]
luminosity, referring the flux ratios to this line. No narrow lines are
visible in 3C~273 and we only report its broad \Ha\ flux. Data for 3C~270 are
taken from \citet{ho97} and corrected for Galactic reddening.
\end{landscape}
\twocolumn

%% file: tab2.tex
\begin{table*}
  \begin{center}
    \caption{Multiwavelength data and spectroscopic classification}
    \label{fulltable2}
    \begin{tabular}{l| c| c c| c c| c| c c c}
      \hline \hline
Name & redshift & \multicolumn{2}{|c|}{ Emission lines}& \multicolumn{2}{|c|}{
  Radio emission}  & Host magnitude & \multicolumn{3}{|c}{Classification}  \\
\hline
     & &  H$\alpha$  & [O~III] & L$_{178}$ &  P$_{core}$ & M$_{H}$ &  FR & spec & Method\\ 
\hline 
3C~015   &  0.073    &   40.40    &  40.60  &   33.30 &  31.64 &  -25.29    &   & LEG  &E.I.\\
3C~017   &  0.2198   &   41.88    &  41.99  &   34.44 &  32.94 &  -24.81$^*$& 2 & BLO  &E.I.\\
3C~018	 &  0.188    &   41.93    &  42.55  &   34.27 &  32.00 &    --      & 2 & BLO  &E.I.\\
3C~020   &  0.174 & 41.37&  41.54 & 34.55 &30.44 &  -24.64$^*$& 2 & HEG & E.I. \\
3C~028	 &  0.1952   &   41.51    &  40.96  &   34.24 &  29.33 &    --      &   & ELEG  &E.I.\\
3C~029	 &  0.0448   &   40.06    &  40.09  &   32.84 &  30.63 &  -25.44    & 1 & LEG  &E.I.\\
3C~031   &  0.0167   &   39.83    &  39.46  &   32.01 &  29.75 &  -25.51    & 1 & LEG  &E.I.\\
3C~033	 &  0.0596   &   41.63    &  42.18  &   33.65 &  30.36 &  -24.75    & 2 & HEG  &E.I.\\
3C~033.1 &  0.1809   &   41.85    &  42.30  &   34.07 &  31.19 &  -24.47$^*$& 2 & BLO  &E.I.\\
3C~035	 &  0.0670   &   40.22    &  40.01  &   33.05 &  30.36 &  -25.17    & 2 & --   &\\
3C~040	 &  0.0185   &   39.08    &  39.22  &   32.29 &  30.66 &    --      & 1 & LEG  &E.I.\\
3C~052	 &  0.2854   & $<$40.64   &  --     &   34.53 &  31.29 &  -26.74$^*$& 2 & --   &\\
3C~061.1 &  0.184    &   42.05    &  42.47  &   34.47 &  30.49 &  -23.50$^*$& 2 & HEG  &E.I.\\
3C~063   &  0.175 & 41.54&  41.63 & 34.21 &31.12 &    --      &   & HEG & E.I. \\
3C~066B	 &  0.0215   &   40.11    &  40.05  &   32.40 &  30.27 &  -26.25$^*$& 1 & LEG  &E.I.\\
3C~075N	 &  0.0232   &   39.58    & $<$39.92&   32.49 &  29.67 &  -24.51$^*$& 1 & --   &\\
3C~076.1 &  0.0324   &   39.89    & $<$39.85&   32.46 &  29.37 &  -24.08$^*$& 1 & --   &\\
3C~078	 &  0.0286   &   39.73    &  39.41  &   32.51 &  31.25 &  -26.16    & 1 & LEG  &D.D.\\
3C~079	 &  0.2559   &   42.39    &  42.86  &   34.78 &  31.39 &  -25.27    & 2 & HEG  &E.I.\\
3C~083.1 &  0.0255   &   39.40    & $<$39.50&   32.57 &  29.48 &  -26.70$^*$& 1 & --   &\\
3C~084	 &  0.0176   &   41.28    &  41.60  &   32.62 &  32.46 &  -25.99    &   & LEG  &E.I.\\
3C~088	 &  0.0302   &   39.98    &  40.14  &   32.49 &  30.57 &  -24.81    & 2 & LEG  &E.I.\\
3C~089   &  0.1386   &   40.28    & $<$40.51&   34.01 &  31.39 &  -26.22    & 1 & --   &\\
3C~093.1 &  0.2430   &   42.35    &  42.67  &   34.24 &   --   &    --      &  & HEG  &E.I.\\
3C~098   &  0.0304   &   40.52    &  41.00  &   32.99 &  29.87 &  -24.38    & 2 & HEG  &E.I.\\
3C~105	 &  0.089    &   40.89    &  41.45  &   33.54 &  30.46 &  -24.31    & 2 & HEG  &E.I.\\
3C~111	 &  0.0485   &    --      &  42.44  &   33.54 &  31.77 &  -25.07    & 2 & BLO  &O.R.\\
3C~123	 &  0.2177   &   41.96    &  42.00  &   35.41 &  32.00 &  -26.58$^*$&   & LEG  &E.I.\\
3C~129	 &  0.0208   &   39.81    & $<$39.85&   32.65 &  29.51 &  -25.11    & 1 & --   &\\
3C~129.1 &  0.0222   &  $<$39.83  &  --     &   32.06 &  28.53 &  -25.61    & 1 & --   &\\
3C~130	 &  0.1090   &  $<$40.17  &  --     &   33.66 &  30.94 &  -28.45    & 1 & --   &\\
3C~132   &  0.214 & 41.37&  41.46 & 34.25 &31.58 &  -26.00    & 2 & LEG & D.D. \\
3C~133	 &  0.2775   &   42.41    &  42.76  &   34.72 &  32.53 &  -25.36$^*$& 2 & HEG  &E.I.\\
3C~135	 &  0.1253   &   41.52    &  42.05  &   33.84 &  30.31 &  -24.47    & 2 & HEG  &E.I.\\
3C~136.1 &  0.064    &   41.41    &  41.44  &   33.13 &  29.16 &  -25.17    & 2 & HEG  &E.I.\\
3C~153   &  0.2769   &   41.60    &  41.63  &   34.56 &  29.94 &  -25.60$^*$& 2 & LEG  &E.I.\\
3C~165	 &  0.2957   &   41.44    &  41.67  &   34.57 &  31.30 &  -25.80$^*$& 2 & LEG  &E.I.\\
3C~166	 &  0.2449   &   41.51    &  41.66  &   34.42 &  32.92 &  -25.32$^*$& 2 & LEG  &E.I.\\
3C~171	 &  0.2384   &   42.45    &  42.89  &   34.51 &  30.55 &  -24.73$^*$& 2 & HEG  &E.I.\\
3C~173.1 &  0.2921   &   41.05    &  40.85  &   34.61 &  31.39 &  -26.48    & 2 & LEG  &O.R.\\
3C~180   &  0.22     &   41.79    &  42.34  &   34.32 &   --   &  -24.94    & 2 & HEG  &E.I.\\
3C~184.1 &  0.1182   &   41.79    &  42.23  &   33.66 &  30.37 &  -24.22$^*$& 2 & BLO  &E.I.\\
3C~192	 &  0.0598   &   40.95    &  41.34  &   33.25 &  29.82 &  -24.68    & 2 & HEG  &E.I.\\
3C~196.1 &  0.198    &   41.56    &  41.52  &   34.31 &  31.82 &  -25.47    & 2 & LEG  &E.I.\\
3C~197.1 &  0.1301   &   40.69    &  40.92  &   33.55 &  30.43 &  -24.94    & 2 & BLO  &E.I.\\
3C~198	 &  0.0815   &   41.31    &  40.97  &   33.19 &   --   &  -23.62$^*$&   & SF   &D.D.\\
3C~213.1 &  0.194    &   41.01    &  41.06  &   33.84 &  31.15 &  -25.02$^*$& 2 & LEG  &E.I.\\
3C~219	 &  0.1744   &   41.55    &  41.77  &   34.53 &  31.69 &  -25.70    & 2 & BLO  &E.I.\\
3C~223	 &  0.1368   &   41.68    &  42.17  &   33.85 &  30.70 &  -24.74    & 2 & HEG  &E.I.\\
3C~223.1 &  0.107    &   41.16    &  41.58  &   33.23 &  30.36 &  -24.95    & 2 & HEG  &E.I.\\
3C~227	 &  0.0861   &   41.08    &  41.75  &   33.74 &  30.58 &  -24.90    & 2 & BLO  &E.I.\\
3C~234   &  0.1848   &   42.64    &  43.11  &   34.47 &  32.04 &  -26.09    & 2 & HEG  &E.I.\\
3C~236	 &  0.1005   &   41.13    &  40.89  &   33.56 &  31.62 &  -25.34    & 2 & LEG  &E.I.\\
3C~258   &  0.165    &   40.96    &  40.19  &   33.85 &   --   &    --      &   & --   &\\
3C~264   &  0.0217   &   39.68    &  39.20  &   32.43 &  30.32 &  -25.09    & 1 & LEG  &E.I.\\
3C~270	 &  0.007 & 39.26&  38.96 & 31.79 &29.57 &  -25.01    & 1 & LEG & E.I. \\
3C~272.1 &  0.0035   &   38.92    &  38.20  &   30.72 &  28.68 &  -24.43    & 1 & LEG  &E.I.\\
3C~273   &  0.1583   &    --      &  --     &   34.62 &  33.65 &    --      &   & BLO  &\\
3C~274	 &  0.0044   &   39.50    &  38.99  &   32.63 &  30.21 &  -25.28    & 1 & LEG  &E.I.\\
\hline                                                                     
  \multicolumn{9}{c}{{Continued on Next Page}} \\                        
    \end{tabular}                                
  \end{center}                                   
\end{table*}                                     
                                                 
\addtocounter{table}{-1}                         
\begin{table*}
  \begin{center}
    \caption{Continued}
    \begin{tabular}{l| c| c c| c c| c| c c c}
      \hline \hline
Name & redshift & \multicolumn{2}{|c|}{ Emission lines}& \multicolumn{2}{|c|}{
  Radio emission}  & Host magnitude & \multicolumn{3}{|c}{Classification}  \\
\hline
     & &  H$\alpha$  & [O~III] & L$_{178}$ &  P$_{core}$ & M$_{H}$ &  FR & Class & Method\\ 
 \hline 
3C~277.3 &  0.0857   &   40.83   &  40.94  & 33.21 &   30.34 &  -24.87    & 2 & HEG  &E.I.\\
3C~284   &  0.2394   &   41.41   &  41.59  & 34.28 &   30.44 &  -25.57    & 2 & HEG  &D.D.\\
3C~285	 &  0.0794   &   40.66   &  40.55  & 33.23 &   30.03 &  -24.53    & 2 & HEG  &E.I.\\
3C~287.1 &  0.2159   &   41.50   &  41.73  & 34.04 &   32.71 &  -25.72    & 2 & BLO  &E.I.\\
3C~288   &  0.246 & 40.86&  40.65 & 34.53 &31.73 &  -26.10$^*$& 2 & LEG & E.I. \\
3C~293   &  0.0450   &   40.18   &  39.80  & 32.77 &   30.67 &  -25.33    &   & LEG  &E.I.\\
3C~296   &  0.0240   &   39.87   &  39.78  & 32.22 &   29.99 &  -26.04    & 1 & LEG  &E.I.\\
3C~300   &  0.27     &   41.78   &  42.01  & 34.60 &   31.27 &  -24.92    & 2 & HEG  &E.I.\\
3C~303	 &  0.141    &   41.33   &  41.74  & 33.77 &   31.94 &  -25.35    & 2 & BLO  &E.I.\\
3C~303.1 &  0.267    &   42.10   &  42.42  & 34.25 &   31.04 &    --      & 2 & HEG  &E.I.\\
3C~305	 &  0.0416   &   40.92   &  41.03  & 32.79 &   30.07 &  -25.26    & 2 & HEG  &E.I.\\
3C~310	 &  0.0535   &   40.32   &  40.05  & 33.56 &   30.72 &  -25.02$^*$& 2 & LEG  &E.I.\\
3C~314.1 &  0.1197   &   40.31   &  39.69  & 33.59 &   29.56 &    --      &   & ELEG  &E.I.\\
3C~315   &  0.1083   &   41.15   &  40.87  & 33.72 &   31.64 &  -24.74$^*$&   & LEG  &E.I.\\
3C~317	 &  0.0345   &   40.35   &  40.35  & 33.12 &   31.02 &  -26.04    &   & LEG  &E.I.\\
3C~318.1 &  0.0453   &   39.95   &  39.36  & 32.72 &   29.12 &  -25.70    &   & --   &\\
3C~319   &  0.192    &   41.16   & $<$40.16& 34.20 &   31.49 &  -24.41$^*$& 2 & --   &\\
3C~321   &  0.096    &   40.50   &  40.91  & 33.49 &   30.89 &  -25.52    & 2 & HEG  &E.I.\\
3C~323.1 &  0.264    &   42.21   &  42.80  & 34.31 &   31.89 &  -26.74    & 2 & BLO  &E.I.\\
3C~326	 &  0.0895   &   40.28   & 40.40   & 33.60 &   30.45 &  -24.33    & 2 & LEG  &O.R.\\
3C~327   &  0.1041   &   41.73   &  42.24  & 33.98 &   30.99 &    --      & 2 & HEG  &E.I.\\
3C~332   &  0.1517   &   41.31   &  41.81  & 33.77 &   30.79 &  -25.38    & 2 & BLO  &E.I.\\ 
3C~338   &  0.0303   &   40.25   &  39.57  & 32.99 &   30.34 &  -26.21$^*$& 1 & LEG  &E.I.\\
3C~346   &  0.161 &   -- &  41.24 & 33.88 &32.18 &  -25.84    & 2 &  -- &      \\
3C~348   &  0.154    &   41.29   &  40.40  & 35.35 &   30.80 &    --      &   & ELEG  &D.D.\\
3C~349   &  0.205 & 41.50&  41.69 & 34.20 &31.35 &  -24.82$^*$& 2 & LEG & E.I. \\
3C~353	 &  0.0304   &   40.42   &  40.14  & 33.69 &   30.61 &  -24.77    & 2 & LEG  &E.I.\\
3C~357   &  0.1662   &   40.92   &  40.95  & 33.86 &   30.63 &  -25.83$^*$& 2 & LEG  &D.D.\\
3C~371   &  0.0500   &   40.94   &  40.94  & 32.33 &   31.85 &  -25.36    &   & LEG  &E.I.\\
3C~379.1 &  0.256    &   41.41   &  41.86  & 34.16 &   30.90 &  -25.69    & 2 & HEG  &E.I.\\
3C~381   &  0.1605   &   41.79   &  42.37  & 34.06 &   30.63 &  -24.81    & 2 & HEG  &D.D.\\
3C~382   &  0.0578   &   41.39   &  41.78  & 33.19 &   31.22 &  -26.03    & 2 & BLO  &E.I.\\
3C~386	 &  0.0170   &   40.17   & $<$40.20& 32.18 &   28.95 &  -24.57    &   & --   &\\
3C~388   &  0.091    &   40.83   &  40.70  & 33.70 &   31.15 &  -26.20    & 2 & LEG  &E.I.\\
3C~390.3 &  0.0561   &   41.57   &  42.08  & 33.54 &   31.46 &  -24.84    & 2 & BLO  &E.I.\\
3C~401   &  0.2010   &   41.01   &  41.05  & 34.38 &   31.67 &  -25.03$^*$& 2 & LEG  &E.I.\\
3C~402	 &  0.0239   &   39.08   & $<$39.42& 32.11 &   29.79 &  -24.77$^*$& 1 & --   &\\
3C~403	 &  0.0590   &   41.20   &  41.75  & 33.16 &   29.96 &  -25.27    & 2 & HEG  &E.I.\\
3C~403.1 &  0.055 & 39.99&  39.86 & 32.98 & --   &  -24.36    &   & LEG & E.I. \\
3C~410   &  0.248 & 41.86&  42.02 & 34.80 &33.43 &    --      & 2 & BLO & E.I. \\
3C~424   &  0.127    &   41.07   &  40.80  & 33.78 &   30.87 &  -23.96$^*$&   & LEG  &E.I.\\
3C~430   &  0.0541   &   40.12   &  40.33  & 33.36 &   30.06 &  -25.28    & 2 & LEG  &O.R.\\
3C~433	 &  0.1016   &   41.40   &  41.67  & 34.16 &   30.11 &  -25.79$^*$&   & HEG  &E.I.\\
3C~436   &  0.2145   &   41.07   &  41.56  & 34.37 &   31.39 &  -25.50    & 2 & HEG  &D.D.\\
3C~438   &  0.290    &   41.55   & $<$41.46& 35.07 &   31.65 &  -26.57    & 1 & --   &\\
3C~442	 &  0.0263   &   39.78   &  39.21  & 32.39 &   28.49 &    --      &   & LEG  &E.I.\\
3C~445   &  0.0562   &    --     &  42.50  & 33.26 &   31.42 &    --      & 2 & BLO  &O.R.\\
3C~449	 &  0.0171   &   39.71   &  39.19  & 31.87 &   29.38 &  -24.80$^*$& 1 & LEG  &E.I.\\
3C~452	 &  0.0811   &   41.16   &  41.34  & 33.94 &   31.34 &  -24.92    & 2 & HEG  &E.I.\\
3C~456   &  0.2330   &   42.48   &  42.81  & 34.23 &   31.57 &    --      & 2 & HEG  &E.I.\\
3C~458   &  0.289 & 41.58&  42.03 & 34.58 &30.88 &    --      & 2 & HEG & E.I. \\
3C~459	 &  0.2199   &   42.17   &  42.03  & 34.55 &   33.20 &  -25.34$^*$& 2 & BLO  &E.I.\\
3C~460   &  0.268    &   42.09   &  41.78  & 34.25 &   31.52 &    --      & 2 & LEG  &E.I.\\
3C~465   &  0.0303   &   40.15   &  39.81  & 32.89 &   30.74 &  -26.44$^*$& 1 & LEG  &E.I.\\
\hline                                                          
    \end{tabular}                                               
  \end{center}                                                  
  Column description: (1) 3CR name; (2) redshift from \citet{spinrad85}; (3) and (4) logarithm
  of H$\alpha$ and [O~III]$\lambda$5007 luminosities [erg s$^{-1}$] from \citet{buttiglione09alias}; (5) radio luminosity at
  178 MHz [erg s$^{-1}$ Hz$^{-1}$] from \citet{spinrad85}; (6) radio core power at 5 GHz [erg
  s$^{-1}$ Hz$^{-1}$] from \citet{baldi09}; (7) host
  H magnitude from 2MASS \citep{skrutskie06} 
(or from HST \citep{donzelli07} for the objects marked with a $^*$);
  (8): morphological FR type; (9) spectroscopic classification into High
  Excitation Galaxy (HEG); 
  Low Excitation Galaxy (LEG); Broad Line Object (BLO); Extremely Low
  Excitation Galaxy (ELEG); (SF) starforming galaxy; (--) unclassified.  Column (10)
  classification method: E.I. - excitation index; D.D. - diagnostic diagrams; O.R.
  - emission line radio correlation. 
\end{table*}